# Spin-Diffusion Lengths in Metals and Alloys, and Spin-Flipping at Metal/Metal Interfaces: an Experimentalist's Critical Review.

Jack Bass and William P. Pratt Jr.
Department of Physics and Astronomy
Michigan State University, East Lansing, MI 48824

## Abstract

In magnetoresistive (MR) studies of magnetic multilayers composed of combinations of ferromagnetic (F) and non-magnetic (N) metals, the magnetic moment (or related 'spin') of each conduction electron plays a crucial role, supplementary to that of its charge. While initial analyses of MR in such multilayers assumed that the direction of the spin of each electron stayed fixed as the electron transited the multilayer, we now know that this is true only in a certain limit. Generally, the spins 'flip' in a distance characteristic of the metal, its purity, and the temperature. They can also flip at F/N or N1/N2 interfaces. In this review we describe how to measure the lengths over which electron moments flip in pure metals and alloys, and the probability of spin-flipping at metallic interfaces. Spin-flipping within metals is described by a spin-diffusion length, $l_{sf}^M$, where the metal M = F or N. Spin-diffusion lengths are the characteristic lengths in the current-perpendicular-to-plane (CPP) and lateral non-local (LNL) geometries that we focus upon in this review. In certain simple cases, $l_{sf}^N$ sets the distance over which the CPP-MR and LNL-MR decrease as the N-layer thickness (CPP-MR) or N-film length (LNL) increases, and $l_{sf}^F$ does the same for increase of the CPP-MR with increasing F-layer thickness. Spin-flipping at M1/M2 interfaces can be described by a parameter, $\delta_{M1/M2}$, which determines the spin-flipping probability, $P = 1 - \exp(-\delta)$. Increasing $\delta_{M1/M2}$ usually decreases the MR. We list measured values of these parameters and discuss the limitations on their determinations.

**Organization**

This review is organized as follows. Section I provides a brief history and overview, defines the lengths of interest, briefly explains the physics underlying the spin-diffusion lengths that are the focus of the review, and discusses caveats on theoretical analysis and limitations on the measurements of the parameters of interest: the transport mean-free-path, $\lambda_t$; the spin-diffusion length in non-magnetic (N) metals or alloys, $l_{sf}^N$; the spin-diffusion length in ferromagnetic (F) metals or alloys, $l_{sf}^F$; and the interfacial spin-flip parameter, $\delta$, where the spin-flipping probability is $P = 1 - \exp(-\delta)$. Section II describes the different ways in which these parameters have been measured, and gives more specifics of their limitations. Section III contains four tables. Table I lists values of $l_{sf}^N$ for well-defined Cu- and Ag-based alloys at 4.2K. $l_{sf}^N$ should be intrinsic to each alloy, and the values are used to test quantitatively both the Valet-Fert (VF) theory of the CPP-MR and two experimental techniques that use it. Table II lists values of $l_{sf}^N$ in nominally pure metals at temperatures T from 4.2K to 293K. $l_{sf}^N$ should be unique to each sample at 4.2K, but should be intrinsic in sufficiently high purity samples at 293K. Table III lists values of $l_{sf}^F$ in ferromagnetic metals and alloys, mostly at 4.2K. Table IV lists values of $\delta_{N1/N2}$ for several N1/N2 metal pairs at 4.2K. Each table is preceded by some comments about the results. Section IV contains a brief summary and our conclusions.

## I. History, Overview, Definitions, and Caveats and Limitations.
### IA. History and Overview.

The discovery in 1988 of Giant Magnetoresistance (GMR) in ferromagnetic/non-magnetic (F/N) metallic multilayers [1] [2] stimulated the growth of a new subfield of transport studies in magnetic materials, now often called Spintronics. In spintronics, the electron's magnetic moment (or spin, which points opposite to the moment) plays a fundamental role supplementary to that of the electron's charge. Preceding the discovery of GMR, pioneering Lateral Non-local (LNL) studies measured [3] the distance over which electron moments (spins) flipped (spin-diffusion length $l_{sf}$) as current passed along a very high purity, annealed Al foil across which two separated F-strips were deposited (see Fig. 12a#2 below). The resulting long spin-diffusion length ($l_{sf}^{Al} \sim 450$ μm at 4K) led to the expectation that such spin-flipping was negligible in GMR multilayers, even though they are much less pure. Thus early analyses of both Current-in-Plane (CIP)-[1] [4] and Current-Perpendicular-to-Plane (CPP)-MRs [5-8] neglected spin-flipping, assuming that the direction of an electron's moment stays fixed as the electron propagates through the multilayer. Even with this assumption, theories of the CIP-MR are relatively complex, in part because the mean-free-paths, λ, for total scattering of electrons (both without and with spin-flipping), are fundamental lengths in the problem [4]. Since usually λ << $l_{sf}$, variations of the CIP-MR with layer thickness are determined mainly by λ. For the CPP-MR [9-11], in contrast, the assumption of no spin-flipping led initially to a simple two-current series resistor (2CSR) model, in which currents of electrons with moments *up* or *down* relative to a fixed direction propagate independently. In the magnetic state where the moments of adjacent F-layers are oriented anti-parallel (AP) to each other, this model gives a total specific resistance (sample area A times resistance R) that is just the sum of effective resistivities (ρ*) times layer thicknesses (t) within the F- and N-layers, plus effective interface specific resistances (AR*) [6] [7, 8]. We will define ρ* and AR*, and present the 2CSR model in more detail in Section IIB. For the moment, we emphasize that the only lengths in this model for the CPP-MR are the thicknesses $t_F$ and $t_N$ of the F- and N-layers; unlike the CIP-MR, λ is generally not a characteristic length in the CPP-MR (but see Appendix C). In 1993, the Valet-Fert (VF) theory [8] of the CPP-MR included effects of finite spin-diffusion lengths in both N- and F-metals ($l_{sf}^N$ and $l_{sf}^F$) in a convenient way that stimulated experimental studies to look for them. CPP-MR studies, first in dilute N-metal alloys [12] and then in the F-alloy Permalloy (Py = $Ni_{1-x}Fe_x$ with x ~ 0.2)[13], found that values of $l_{sf}^N$ or $l_{sf}^F$ were sometimes small enough to be comparable to experimentally interesting layer thicknesses, $t_N$ or $t_F$. Inserting a finite $l_{sf}^N$ or $l_{sf}^F$ affects the magnitude of the CPP-MR, usually reducing it from what it would have been with infinite $l_{sf}^N$ or $l_{sf}^F$, but occasionally enhancing it (see [14, 15] and Figs. 5 and 6) While spin-flipping within N- and F-metals is now regularly taken into account, most studies still assume that electrons do not flip their spins when they cross interfaces. We will argue that this assumption can be wrong, even for simple N1/N2 interfaces. Spin-flipping at interfaces is especially interesting at F/S (S = superconductor) and F/N interfaces, where, unfortunately, little reliable information is yet available. For F/N interfaces, we'll discuss in Section II.B2.b2.b what little is known. Interestingly, inclusion of modest spin-flipping at F/N interfaces does not greatly affect the CPP-MR of simple [F/N]$_\mathcal{N}$ multilayers [16], where $\mathcal{N}$ is a significant number of bilayers. But it does affect the CPP-MR of symmetric exchange-biased spin-valves (EBSVs) [16], made up of only two equal-thickness F-layers, the magnetization of one pinned in a fixed direction by an adjacent antiferromagnet (AF), and the magnetization of the other free to reverse from parallel (P) to antiparallel (AP) to that of the first [17]. In proximity-effect F/S systems, effects of spin-flipping in the bulk F-metal have been seen in damped oscillatory behavior of the superconducting correlations, in agreement with predictions [18-20]. Expected longer-range penetration into F of triplet-state superconducting correlations is also predicted to be attenuated by spin flipping in the bulk of F [19, 21], and the triplet correlations themselves may be affected by spin flipping in the bulk of S and at the F/S interfaces. But these other predictions have yet to be confirmed.

In this review, we examine, from an experimentalist's perspective, what we believe has been learned about $l_{sf}^N$, $l_{sf}^F$, and $\delta_{N1/N2}$, and also some less clearcut studies of $\delta_{F/N}$. We have organized the review for readers with different levels of interest. Those interested only in the results can read just section I--which briefly outlines the lengths and physics involved, and describes the caveats and limitations of the various types of measurements, and section III--which contains tables of the published values of $l_{sf}^N$, $l_{sf}^F$, and $\delta_{N1/N2}$, along with our comments. Values of $l_{sf}^N$ are obtained from CPP-MR, Lateral Non-Local (LNL), and some Weak Localization (WL) measurements. Values of $l_{sf}^F$ are obtained mostly from CPP-MR, with a few values from LNL measurements. Values of $\delta_{N1/N2}$ have so far been obtained only by CPP-MR. For those interested in details, Section II describes the ways used to derive $l_{sf}^N$, $l_{sf}^F$ and $\delta_{N1/N2}$, and gives more specifics of limitations, and Appendices A-C discuss in more detail important parameters and issues, and respond to a critique of the fundamental assumptions underlying this review.

WL, Conduction Electron Spin-Resonance (CESR), and Superconducting Tunneling measurements have all been used to derive spin-relaxation times, $\tau_{sf}$, in metals and alloys. We include only studies of $\tau_{sf}$ where the authors explicitly calculated $l_{sf}$, referring readers interested in $\tau_{sf}$ to sources such as [22] [23, 24].

Because of both experimental and theoretical uncertainties, it is important to compare values for nominally identical parameters determined in different ways. Making such comparisons possible is one of the tasks of this review. We sometimes have clear views as to which of conflicting analyses and derived values are most reliable. We explain our reasoning, but warn that others will not necessarily agree with us.

**IB. Lengths and 'Physics'.**

To analyze F/N multilayer structures, we must distinguish several different lengths.

In an isolated N-metal, there are three: the transport (momentum exchange) mean-free-path, $\lambda_t$; the spin-flip length, $\lambda_{sf}$; and the spin-diffusion length, $l_{sf}$. Qualitatively, electrons are pictured as moving ballistically between collisions, but making many collisions as they traverse a sample (diffusive transport). $\lambda_t$ is the mean distance between collisions of all kinds (both spin-direction conserving and spin-flipping). Defining a mean-time $\tau$ between collisions, gives $\lambda_t = v_F\tau$, where $v_F$ is the Fermi velocity. Similarly, $\lambda_{sf}$ is the mean distance between spin-flipping collisions. Defining $\tau_{sf}$ as the mean-time between spin-flipping events gives $\lambda_{sf} = v_F\tau_{sf}$. $l_{sf}$, in contrast, is the mean distance that electrons diffuse between spin-flipping collisions (Appendix A shows that $l_{sf}$ obeys a diffusion equation). Except for a numerical factor, it is the geometric mean of $\lambda_t$ and $\lambda_{sf}$. In a single N-metal, the standard form is [8]:

$$l_{sf} = \sqrt{D\tau_{sf}} = \sqrt{(1/3)\lambda_t v_F \tau_{sf}} = \sqrt{(1/3)\lambda_t \lambda_{sf}}, \tag{1}$$

where D is the 'diffusion constant', and the usual ordering of lengths is $\lambda_t < l_{sf} < \lambda_{sf}$.

In an isolated F-metal, one must consider separate propagation of electrons with moments *up* or *down*. Asymmetric scattering of majority (electron magnetic moment along ($\uparrow$) the local F-layer moment) and minority (electron moment opposite to ($\downarrow$) that of the local F-layer) electrons leads to separate mean-free-paths, $\lambda_\uparrow^F$ and $\lambda_\downarrow^F$, with scattering of ($\downarrow$) electrons usually stronger. This asymmetric scattering also leads to separate spin-diffusion lengths, $l_\uparrow^F$ and $l_\downarrow^F$, but in the equations of primary interest, these combine into a single 'spin-diffusion' length, $l_{sf}^F$, according to Appendix A $[1/l_{sf}^F]^2 = [1/l_\uparrow^F]^2 + [1/l_\downarrow^F]^2$ [8, 25]. We show in Appendix A that applying Eq. 1 to each spin channel within VF theory, and using the relation for $l_{sf}^F$ just given, produces an equation for $l_{sf}^F$ in which the $\lambda$ in Eq. (1) is not $\lambda_t$, but a new quantity $\lambda^* = \lambda_t(1-\beta_F^2)$ -- $\beta_F$ is defined in Appendix A and Section.II.B1, and the fraction under the square root for both $l_{sf}^N$ and $l_{sf}^F$ is not (1/3), but (1/6). Thus we obtain:

F-metal. $$l_{sf}^F = [\lambda^* \lambda_{sf}^F /6]^{1/2} = [(1-\beta_F^2)\lambda_t^F \lambda_{sf}^F /6]^{1/2}. \tag{2}$$

For an N-metal, $\beta_F = 0$, and Eq. (2) reduces to just

N-metal. $$l_{sf}^N = \sqrt{(\lambda_t^N \lambda_{sf}^N)/6} \tag{3}$$

We assume from here on, that Eqs. (2) and (3) apply to the F and N components of general F/N multilayers.

In most samples of nominally 'pure' metals, electrons are scattered at cryogenic temperatures mostly by an unknown combination of (generally) unknown impurities, and at room temperature by a combination of these impurities and phonons. In sputtered or evaporated samples, the scattering from these impurities usually remains important to above room temperature. In such a case, any experimentally derived value of $l_{sf}^N$ strictly represents only a property of the given sample being measured. To obtain an intrinsic value of $\lambda_{sf}^N$, one must have either a sample of high enough purity that phonon scattering is dominant at the measuring temperature, or an alloy in which a known concentration c of a known impurity is dominant.

If a dominant impurity in a host has no local magnetic moment, it flips electron spins by spin-orbit scattering. If the spin-orbit cross-section, $\sigma_{so}$, for this impurity in this host is known from CESR (see, e.g., [26]), then $\lambda_{sf}^N$ is given by:

$$\lambda_{sf}^N = \lambda_{so}^N = [1/(nc\sigma_{so})], \tag{4}$$

where n is the number of host atoms per unit volume. CESR values of $\sigma_{so}$ are given for a number of Cu-based alloys (and some Ag-, and Al-based ones) in [26]. For such a dilute, known, impurity concentration, both $\lambda_t^N$ and $\lambda_{sf}^N$ are proportional to (1/c), giving $l_{sf}^N \propto \lambda_t^N$. If $\lambda_t^N$ can be determined (see Appendix B), then $l_{sf}^N$ can be calculated from Eq. (3) and compared with experiment. In Table I we compare experimental and calculated values of $l_{sf}^N$ at 4.2K for several dilute alloys in which spin-orbit scattering is dominant. We take the observed agreement between these values as evidence that both the VF theory and the experimental techniques used are valid, and, thus, that little if any 'mean-free-path' effects (see Appendix C) are needed. Some of the impurities included in Table I have a local moment, in which case spin-flipping is produced by spin-spin scattering. Estimating $\lambda_{sf}^N$ is then more complicated, and we refer the interested reader to Ref. [27].

Any additional source of scattering that increases the resistivity ρ will decrease $l_{sf}$ at least as the square root of the inverse of ρ, because the mean-free-path $\lambda_t$ appears under the square root for $l_{sf}$ in both Eqs. (2) and (3), and because Eq. (5) below and Appendix B show that $\lambda_t$ is inversely proportional to ρ. If the source also flips spins, it may decrease $l_{sf}$ inversely with ρ. We will discuss this topic further in section IC, and test for an inverse relationship between $l_{sf}$ and ρ, in Tables II and III by including a column of the product $\rho l_{sf}$, and in Figs. 14-16 by plotting $l_{sf}$ vs 1/ρ.

To clarify how finite $l_{sf}^N$ and $l_{sf}^F$ affect the MR of a simple N1/F1/N2/F2 multilayer, we discuss two different ways of looking at the MR, each starting from a simple N1/F1/N2 trilayer, and adding an F2 layer to the right of N2 as a 'detector'.

The first way is to say that asymmetric scattering within F1 causes emerging electrons to be 'spin-polarized' (more precisely, magnetized—i.e. usually more moment ↑ (i.e. along the moment of F1) than moment ↓ (opposite to the moment of F1), and that spin-polarization can be detected, as with polarized light, by putting a 'detector' at the 'end' of N2. If the detector is another fully magnetized F-metal, F2, then there should be a change in the voltage, ΔV, or the related resistance (ΔR = ΔV /I), across the sample when the moment of F2 is reversed from parallel (P) to anti-parallel (AP) to that of F1. The limit where the current arriving at F2 is 'unpolarized' should give ΔR = 0. When ΔR ≠ 0, it should decay exponentially with the separation $L$ (= Δz) between F1 and F2 (see Appendix A and the following).

The second is to say that, as more ↑ electrons than ↓ ones pass through F1, a pileup of excess ↓ electrons must occur on the N1 side of the N1/F1 interface. The system must adjust itself so that, at steady state, the excess ↓ electrons that arrive at the interface 'diffuse away' as fast as they arrive. At the interface, there will be a non-zero difference in chemical potential Δμ = μ↑ - μ↓ --i.e., the number of electrons at the Fermi surface will be larger for whichever of μ↑ or μ↓ is larger. In a free-electon model, |Δμ| = $2\mu_o$ |ΔM| /(3n$\mu_B$) is related to an out of equilibrium magnetization, ΔM, where n is the electron density, $\mu_B$ is the Bohr Magneton, and $\mu_o$ is the magnetic permeability of empty space. This excess local magnetization (or 'spin') is called a 'spin-accumulation' [8]. As shown in Appendix A, Δμ is governed by a diffusion equation with length scale $l_{sf}^N$ or $l_{sf}^F$. In 1D, the solution to this equation has the form Δμ = Aexp($z/l_{sf}$) + Bexp(-$z/l_{sf}$), where the coefficients A and B are determined by boundary conditions. In a simple N/F1(t)/N trilayer, where t is the F1 layer thickness, |Δμ| grows exponentially in N as F1 is approached from the left, may vary or not within F1 depending upon the ratio t/$l_{sf}^{F1}$, and then decays exponentially in N2 with increasing z. The experimental procedures used to determine $l_{sf}^N$ are normally arranged so that the spin-accumulation decays exponentially in N2 away from the F1/N2 interface. More precisely, the contribution of spin-accumulation to the voltage (resistance) across the sample changes as the moment of F2 is reversed, and it is usually the exponential decay of this change with separation ($L$ = Δz) between F1 and F2 that is used to determine $l_{sf}^N$. The procedures used to determine $l_{sf}^F$ are somewhat more complex, generally requiring a solution of the VF equations with appropriate boundary conditions, including possible changes in μ at the interfaces (neglected for simplicity in this simple discussion). In LNL measurements, the net flow of applied current, and the decay of spin-accumulation, occur in different parts of the sample (hence the appellation 'non-local').

With Δμ = μ↑ - μ↓ defined, we can specify more precisely the effect of finite $l_{sf}$. As noted above, and shown in Appendix A, $l_{sf}^N$ is the characteristic length over which Δμ varies within an N-metal, and $l_{sf}^F$ is the length over which it varies within an F-metal [8]. In carefully chosen geometries, the CPP-MR (see, e.g., Fig. 9 below) and LNL-MR (Fig. 13 below) can decrease exponentially with increasing N-layer thickness, $t_N$, on a scale set by $l_{sf}^N$, and the CPP-MR can increase with increasing F-layer thickness, $t_F$, on a scale set by $l_{sf}^F$ (see, e.g., Figs. 5 and 6 below)

### IC. Caveats and Limitations on Measurements.

Most published measurements of $l_{sf}^N$, $l_{sf}^F$ are at 4.2K or near room temperature (RT ≈ 293K).

At 4.2K, scattering by magnons or phonons is negligible, and $l_{sf}^N$ and $l_{sf}^F$ are determined by spin-orbit or spin-spin scattering from defects or impurities [27]. As noted above, values of $l_{sf}^N$ or $l_{sf}^F$ at 4.2K are, thus, intrinsic only for binary alloys where scattering from a known concentration of a known impurity dominates. At 293K, in contrast, scattering by phonons can dominate the resistivity of pure enough metals, in which case $l_{sf}^N$ or $l_{sf}^F$ should be intrinsic. As also noted above, in sputtered or evaporated N- or F-metals, scattering from residual defects and impurities can be comparable to that from phonons at 293K, in which case $l_{sf}^N$ or $l_{sf}^F$ would not be intrinsic to the host metal. Moreover, in F-metals scattering by magnons increases with increasing temperature [27], as does the generalization of spin-orbit scattering in all metals [28]. Along with phonon scattering, these reduce $l_{sf}$ as temperature T increases, and different combinations could affect correlations of $l_{sf}^N$ or $l_{sf}^F$ with 1/ρ.

For N1/N2 interfaces, one can be sure that $\delta_{N1/N2}$ is fundamental only for a given interfacial structure. It isn't known how sensitive δ might be to interface intermixing. Perhaps it is not, since calculated interface specific resistances, 2AR, are often not sensitive to details of interfacial intermixing [29-31]. So far, only one technique has been used to measure $\delta_{N1/N2}$, and no calculations of $\delta_{N1/N2}$ have been made; thus how intrinsic the results are is not clear. Most models used to analyze experimental data assume identical free-electron Fermi surfaces in both the N- and F-metals. In this case, the mean-free-paths, λ, in both the N- and F-metals are not characteristic lengths in the CPP-MR, and play no direct role [6, 8]. If there is also no spin-flipping, one obtains the 2CSR model. In the past few years, several theorists have shown [32-36] that taking account of real Fermi surfaces might cause the interface specific resistances to change as the layer thicknesses vary from much larger to much smaller than the mean-free-paths in the layers. Such changes in interface specific resistance with layer thickness are called 'mean-free-path' (mfp) effects. For thick enough layers, these models provide a more rigorous justification for use of the 2CSR model when spin-flipping is absent. Said another way, in this limit the 2CSR model provides a convenient way to parameterize the experimental data in terms of layer resistivities and interface specific resistances, leaving the detailed understanding of these parameters to be handled separately. For thinner layers, these calculations mean that the situation is less clear, and experiments must be examined carefully. At one extreme, it is argued [32] that mfp effects might be the source of the phenomena that we attribute to finite spin-diffusion lengths. Since this argument calls into question the basis of the present review, we must address it in detail. To avoid a major diversion in the body of the review, we do so in Appendix C. Our conclusion is closer to the other extreme, that, so far, any mfp effects appear to be modest ($\leq 10\%$), almost always falling within experimental uncertainties, We argue in Appendix C that, with one possible exception, there is no compelling evidence that mfp-effects cause significant deviations from the free-electron VF equations, and substantial experimental evidence to the contrary.

The Valet-Fert (VF) equations used to derive $l_{sf}^N$, $l_{sf}^F$, and $\delta_{N1/N2}$ from CPP-MR data are strictly only the lowest order expansion in the ratio(s) $\lambda/l_{sf}$, and in some alloys this ratio is not $<<1$. Penn and Stiles [37] recently showed numerically that they remain good approximations even when this ratio is close to one.

Most models used to analyze both CPP-MR and LNL data are one-dimensional (1D)—i.e. they assume that a constant current density flows uniformly through the sample. We'll see that some samples and geometries satisfy the conditions needed for this to be true, but that others do not. Parameters determined by those that do not are at least somewhat suspect.

To obtain the most reliable values of $l_{sf}^N$ with Lateral Non-Local (LNL) measurements and low resistance, metallic F/N contacts, the sample width, $W$, should be much less than the sample length, $L$ (i.e., $W << L$). LNL measurements with tunneling contacts don't suffer this limitation. We'll also argue that, for low resistance, metallic F/N contacts, different equations must be used to analyze data when the F/N interface resistance is less than, in-between, or larger than the effective resistances of the thin F and N films (see Eqs. 16a-c below). If so, some investigators used inappropriate equations to analyze their data. For LNL measurements with tunneling contacts, Eq. 16c should always be valid.

## II.  Determining $\lambda_t$, $l_{sf}^N$, $l_{sf}^F$, and $\delta_{N1/N2}$.

In section IIA we explain how to determine the transport mean-free-path, $\lambda_t$. Section IIB contains background information on the CPP-MR and then details of how it is used to determine $l_{sf}^N$, $l_{sf}^F$, and $\delta_{N1/N2}$. In Section IIB2,b2.b, we also examine some inferences about $\delta_{F/N}$. Section IIC describes how Lateral Non-local (LNL) measurements are used to determine $l_{sf}^N$ and $l_{sf}^F$. Section IID briefly outlines the Weak-Localization (WL) technique used to determine the spin-diffusion length limited by spin-orbit scattering, $l_{so}^N$.

Finding $l_{sf}^N$, $l_{sf}^F$, or $\delta_{N1/N2}$ from the CPP-MR or LNL measurements involves measuring the change in specific resistance, AΔR (for CPP-MR), or just the change in resistance, $\Delta R = R(AP) - R(P)$ (for LNL), when the magnetizations of two F-layers are switched by an external magnetic field H from parallel (P) to anti-parallel (AP) to each other. Determining AR(AP), AΔR, and ΔR, thus requires the ability to achieve both P and AP states. The P state can be obtained simply by applying a field H large enough to align all F-layers parallel to the field. The AP state is harder to produce, but has been achieved in several ways. Firstly, some simple [F/N]$_N$ multilayers adopt an AP ordering of the F-layer magnetizations, either in their as-prepared state [38] (Fig. 1a), or because the magnetic exchange coupling between adjacent F-layers is antiferromagnetic (AF) [1, 2]. Secondly, in a multilayer of the form [F1/N/F2/N]$_N$, the two F-layers can have different switching fields if they are different metals or alloys, or if they have different layer thicknesses and/or widths (this last is used especially in LNL studies). Fig. 1b illustrates the resulting AR$_T$(H). Thirdly, the magnetization direction of one of the F-layers can be 'exchange-bias pinned' [17] to an adjacent AF-layer, and the other 'free' F-layer placed so far away that exchange coupling is negligible. The free layer in such an exchange-biased spin-valve (EBSV) then switches reversibly at much lower values of H than needed to 'unpin' the pinned layer. Fig. 1c shows the 'minor loop' for an EBSV, where the pinned layer stays pinned.

### IIA. Finding $\lambda_t$.

Appendix B explains how one defines $\lambda_t$ for a given metal. Since the mean-free-paths of electrons may vary over the Fermi surface, $\lambda_t$ must be an average over this surface. Traditionally $\lambda_t$ is estimated from the relation ([39] and appendix B)

$$\lambda_t = \rho_b l_b / \rho_t, \qquad (5)$$

where the product $\rho_b l_b \sim 1$ f$\Omega$m$^2$ is assumed to be a temperature independent constant for given host-metal and $\rho_t$ is the sample resistivity at temperature T. The constant $\rho_b l_b$ can be calculated (assuming free-electrons [39] or real Fermi surfaces [40]), or measured from size-effect or anomalous skin-effect studies [40]. We argue in Appendix B that the uncertainty for the most widely studied metals, Cu and Ag, probably doesn't exceed 50%.

Determining $\lambda_t$ generally starts with a four-probe, CIP measurement of the electrical resistivity, $\rho(T)$, of a thin film, often using the van der Paaw technique [41]. For LNL or WL studies, this may be the thin sample film itself, in which case the resistivity can include a component due to surface scattering. For samples in the CPP-geometry, it is usually a CIP measurement of a separate film prepared in the same way as the CPP sample film, and made several times thicker than the expected mean-free-path, to minimize any surface contribution.

### IIB. CPP-MR.
### IIB.1. Background

As noted in section I, if spin-flipping is negligible, the CPP-MR can often be well described by a simple 2CSR model. In this model, currents for *up* and *down* electrons propagate independently and in parallel, and AR for each current is just the sum of appropriate resistivities ($\rho_F^\uparrow$ or $\rho_F^\downarrow$) times F-layer thickness $t_F$ in the F-layers, $2\rho_N t_N$ (because only one spin direction is involved) in the N-metal, and $AR_{F/N}^\uparrow$ or $AR_{F/N}^\downarrow$ at each F/N interface. As in section IB, $\uparrow$ and $\downarrow$ mean that the electron moment is oriented along or opposite to the moment of the F-metal through which it is passing. A set of four alternative parameters more convenient for analysing the CPP-MR are: $\rho_F^* = (\rho_F^\downarrow + \rho_F^\uparrow)/4$; $\beta_F = (\rho_F^\downarrow - \rho_F^\uparrow)/(\rho_F^\downarrow + \rho_F^\uparrow)$ for bulk F (see Appendix A); and $AR_{F/N}^* = (AR_{F/N}^\downarrow + AR_{F/N}^\uparrow)/4$; and $\gamma_{F/N} = (AR_{F/N}^\downarrow - AR_{F/N}^\uparrow)/(AR_{F/N}^\downarrow + AR_{F/N}^\uparrow)$ for F/N interfaces. By measuring the F-metal resistivity, $\rho_F = \rho_F^*(1-\beta_F^2)$ ([8] and Appendix A), and the additional N-metal resistivity, $\rho_N$, on separately prepared thin films, the number of unknown parameters can be reduced from five to three. Three is few enough that one can test the applicability of the 2CSR model [7, 42] when spin-flipping is weak, and look for effects of finite $l_{sf}^N$, $l_{sf}^F$, or $\delta_{N1/N2}$, when spin-flipping is stronger.

Valet and Fert (VF) [8] showed how to extend the 2CSR model to include the two new parameters, $l_{sf}^N$ and $l_{sf}^F$. Because their general equations are complex, we write them down only in certain limiting cases, more generally merely noting that they are usually fit to a given set of data numerically, treating $l_{sf}^N$ or $l_{sf}^F$ as the only one, or one of only a few, unknown(s). The VF equations are derived assuming that both F- and N-metals have free-electron Fermi surfaces, with the only difference being the scattering within them. As noted above, including real Fermi surfaces might lead to deviations from the VF equations, or might just modify the values of the parameters, leaving the VF equations essentially intact. To not break the flow of the review, this issue is addressed in Appendix C.

### IIB2 Finding $l_{sf}^N$, $l_{sf}^F$, and $\delta_{N1/N2}$ from **CPP-MR.**

In addition to the different ways of controlling the magnetization orientations of the F-layers described above, different geometries have been used to isolate spin-flipping parameters. To avoid having to continually respecify details of sample geometry and control of magnetic order, we define here acronyms for the CPP-MR, based upon the geometries of Fig. 2. Fig. 2a is a short-wide ($L \ll W$) sample using superconducting (S) cross-strips (CPP-S). This geometry is used with either simple [F/N]$_N$ multilayers (CPP-S/ML) or AF/F/N/F spin-valves (CPP-S/SV). It is limited to low temperatures (so far, only to 4.2K), and has been used only for spin-diffusion lengths shorter than about 100 nm. Fig. 2b is a long, thin ($W \ll L$) CPP-NanoWire Multilayer (CPP-NW/ML), which can be electrodeposited into a cylindrical hole in a polycarbon or Al$_2$O$_3$ substrate. Fig. 2c is a CPP-nanopillar (CPP-NP) with $W \sim L$. These are mostly produced by electron-beam lithography and subtractive ion etching. The last two techniques can be used at room temperature and for longer spin-diffusion lengths.

Since the VF equations are 1D, we must ask whether the CPP-current flows uniformly through samples having the three geometries just described. It does for short ($L \leq 1$ μm), wide ($W \sim 1.2$ mm) CPP samples with superconducting (S) cross-strip leads--CPP-S [42] (Fig. 2a), and for long ($L \sim$ μm), narrow ($W \sim 50$ nm) CPP-nanowires (CPP-NW) [43] (Fig. 2b), the former because the crossed-S strips are equipotentials and the latter because $L >> W$. It does not strictly do so for typical nanopillars (CPP-NP) where $L \sim W$ (Fig. 2c), but becomes better the smaller the sheet resistance, $\rho/t$, of the extended-width N-leads compared to the resistance R of the nanopillar.

### IIB2.a. CPP-S/ML used to determine $l_{sf}^N$ for alloys at 4.2K.

The first CPP-MR determinations of $l_{sf}^N$ involved N = Cu- or Ag-based alloys and application of the 2CSR model to [F/N]$_\mathcal{N}$ CPP-S/ML multilayers with fixed $t_F$ [12]. Including a specific resistance AR$_{S/F}$ for each S/F interface at the ends of the sample [7, 42], and neglecting the difference between $\mathcal{N}$ and $\mathcal{N}$+1, the 2CSR model gives

$$AR_T(AP) = 2AR_{S/F} + \mathcal{N}[\rho_N t_N + \rho_F^* t_F + 2\,AR_{F/N}^*] \quad (6)$$

and

$$A\Delta R = \mathcal{N}^2[\beta_F \rho_F^* t_F + 2\gamma_{F/N} AR_{F/N}^*\,]^2/AR_T(AP). \quad (7a)$$

For use below, we rewrite Eq. (7a) in the form

$$\sqrt{A\Delta R (AR_T(AP))} = \mathcal{N}[\beta_F \rho_F^* t_F + 2\gamma_{F/N} AR_{F/N}^*] \quad (7b)$$

For a set of multilayers with fixed $t_F$, the bracketed quantity on the right-hand-side (RHS) of Eq. 7b is constant, independent of both $\rho_N$ and $t_N$. Eq. 7b then says that a plot of experimental data for the square-root on its LHS versus $\mathcal{N}$ should yield a straight line passing through the origin, and the slope of this line should be independent of $\rho_N$. If we replace a relatively-pure N metal, having a low value of $\rho_N$, by an alloy N' having a large $\rho_{N'}$, the data for N' should fall on the same line as that for N.

Underlying Eqns. 6 and 7 are requirements that $l_{sf}^F >> t_F$ and $l_{sf}^N >> t_N$. In the experiments we describe, the F-metal was Co, and $t_{Co} = 6$ nm was fixed at a value well below the $l_{sf}^{Co}$ listed in Table III. So finite $l_{sf}^F$ is presumably not a problem. In addition, the total thickness, $t_T = \mathcal{N}(t_N + 6)$ was held fixed at either 360 or 720 nm. Decreasing $\mathcal{N}$, thus, requires an increase in $t_N$, and finite $l_{sf}^N$ can become important. For nominally pure Cu or Ag at 4.2K, Table II shows that $l_{sf}^N \geq 200$ nm, long enough that Eq. 7b should apply. If, however, alloying reduces $l_{sf}^N$, then deviations from the straight line predicted in Eq. 7b should be expected, with the fractional deviations increasing with decreasing $\mathcal{N}$. $l_{sf}^N$ is found by analyzing these deviations with the VF equations.

Fig. 1a illustrates the problem of determining AR$_T$(AP) and A$\Delta$R with simple [F/N]$_\mathcal{N}$ multilayers. AR$_T$(P) can be determined simply by increasing the applied magnetic field H until AR$_T$ saturates at its minimum value at high H. However, the data of Fig. 1a provide two potential possibilities for AR$_T$(AP), the as-prepared value of AR$_T$, AR$_T$(0), before any field was applied, or the largest value, AR$_T$(peak), after saturation was achieved. Since AR$_T$(AP) should be the maximum value of AR$_T$, the values of $l_{sf}^N$ given in Table I were determined assuming AR$_T$(AP) = AR$_T$(0). Subsequent studies [38] showed that the as-prepared state of [Co/Cu]$_\mathcal{N}$ and [Co/Ag]$_\mathcal{N}$ multilayers with fixed $t_{Co} \sim 6$ nm often closely approximates the AP state. In addition, systematic use of AR$_T$(Peak) for both pure and alloyed samples gave closely the same values of $l_{sf}^N$ [44]. These values, thus, appear to be reliable to ~20%.

Fig. 3 [12] shows $\sqrt{A\Delta R(AR_T(0))}$ vs $\mathcal{N}$ for pure Ag, AgSn, AgPt, and AgMn alloys, and Fig. 4 [12, 45] shows similar data for Cu and Cu-based alloys. The residual resistivities, $\rho_o$, for the alloys are given in Table I. Note especially that the $\rho_o$ values for AgSn and CuGe are larger than those for AgPt or CuPt. We use this fact in Appendix C as an argument against the importance of 'mean-free-path' effects on these data.

Because Sn is close to Ag in atomic number, and Ge is close to Cu, we expect the spin-orbit cross-sections in both to be small, and indeed the data for AgSn in Fig. 3 and for CuGe in Fig. 4 fall closely along the straight lines through the origin set by the data for nominally pure Ag and Cu. In contrast, the heavy metal Pt has a large spin-orbit cross-section in both Ag and Cu [26]. The data for AgPt and CuPt fall well below that for Ag and Cu and, as shown in Table I, VF fits to the data for $l_{sf}$ in each case agree with values calculated from the spin-orbit cross-sections. Table I shows that similar agreement is found for Cu(Ni) alloys. Because Mn in Ag or Co has a local moment, scattering from it is dominated by spin-spin interactions. The values of $l_{sf}^N$ for Mn found from VF theory are compared in Table I with calculations for spin-spin flipping [27]. Experiments and calculations again agree.

Finally, as noted in section IIB2 above, the AP state can be achieved more certainly using EBSVs. Table I contains two examples, Ag(6%Pt) and Cu(22.7%Ni), of values of $l_{sf}^N$ obtained using EBSVs as described in section IIB2.b2. The good agreement of these values with those for multilayers supports the validity of both techniques.

### IIB2.b. CPP-S/SV used to determine $l_{sf}^F$, $l_{sf}^N$, and $\delta_{N1/N2}$ at 4.2K.

To reliably produce $AR_T(AP)$, subsequent studies of $l_{sf}^N$, $\delta_{N1/N2}$, and $l_{sf}^F$ at 4.2K shifted to EBSVs. Most EBSV-based determinations of $l_{sf}^N$ and $\delta_{N1/N2}$ used F = Py (Py = Permalloy = $Ni_{1-x}Fe_x$ with x ~ 0.2), because the free Py layer flips in fields small enough (~ 20 Oe) that the pinned layer stays well pinned. For measuring $l_{sf}^N$ and $\delta_{N1/N2}$, both Py layers were also taken to be much thicker (typically 24 nm) than $l_{sf}^{Py}$ ~ 5.5 nm, so as to make the free Py-layer flip at a low field and to minimize variations in AΔR due to fluctuations in $t_{Py}$ and in resistances outside the Py layers (see Eq. 10 below). We discuss first the geometry and analysis used for $l_{sf}^F$, and then the common geometry and analysis used for $l_{sf}^N$ and $\delta_{N1/N2}$.

### IIB2.b1. Determining $l_{sf}^F$ with CPP-SVs.

The basic geometry used to determine $l_{sf}^F$ is a symmetric CPP-SV of the form AF/F/N/F, using FeMn as the AF pinning layer, and maintaining equal thicknesses $t_F$ of the two F-layers. Since $t_F$ must be varied over a large range, and pinning effectiveness decreases with increasing $t_F$, care must be taken that AP states are still achieved for the thickest layers. If all spin-diffusion lengths in the sample are long, $l_{sf}^N \gg t_N$ and $l_{sf}^F \gg t_F$, the 2CSR model now gives

$$A\Delta R = 4[\beta_F \rho_F^* t_F + \gamma_{F/N} AR_{F/N}^*]^2 / AR_T(AP), \tag{8}$$

where 
$$AR_T(AP) = AR_{S/F} + AR_{S/AF} + \rho_{AF} t_{AF} + AR_{AF/F} + 2\rho_F^* t_F + 2 AR_{F/N}^* + \rho_N t_N. \tag{9}$$

Note the different treatments of the S/F boundaries next to F and to AF. Since $t_F$ is squared in the numerator, but only linear in the denominator, for large $t_F$, AΔR increases approximately linearly with $t_F$, as shown by the dashed curves in Figs. 5 [13] and 6 [46], where we plot AΔR vs $t_F$ for Py, CoFe, and Co.

If, instead, $l_{sf}^N$ is still long, but $t_F \gg l_{sf}^F$, we must use the more general VF model, and the $t_F$ in the numerator of Eq. 8 is replaced by $l_{sf}^F$ and the denominator reduces to the total AR for just the central 'active' region of the EBSV, lying within $l_{sf}^F$ of each of the two F/N interfaces [11] [13]

$$A\Delta R = 4[\beta_F \rho_F^* l_{sf}^F + \gamma_{F/N} AR_{F/N}^*]^2 / (2\rho_F^* l_{sf}^F + 2 AR_{F/N}^* + \rho_N t_N). \tag{10}$$

In this case, AΔR is constant, independent of $t_F$—i.e., AΔR saturates for large $t_F$.

The signature of a finite $l_{sf}^F$ is, thus, an initial approximately linear growth in AΔR, followed by eventual saturation to a constant value. At $t_F$ between the linear and saturation regimes, AΔR is given by a complex VF expression that must be solved numerically and fit to the data with $l_{sf}^F$ as a fitting parameter. Such fits are shown as solid curves for Py in Fig. 5 and Co(9%Fe) in Fig. 6, with the resulting values of $l_{sf}^F$ listed in Table III. Note that the values of $l_{sf}^F$ are much smaller than the $t_F$ at which AΔR saturates. Rather, they lie close to where AΔR deviates from the 2CSR model dashed lines. A simplified VF picture of why this happens is as follows. The numerator of Eq. (8) reaches its maximum value when $t_F \sim l_{sf}^F$, after which the $t_F$ of Eq. (8) is replaced by the $l_{sf}^F$ of Eq. (10). The denominator, in contrast, starts to decrease even before $t_F = l_{sf}^F$, as the contributions to it from the layers and interfaces outside of the 'active' region begin to disappear. This decrease continues until $t_F \gg l_{sf}^F$, when the denominator becomes constant as in Eq. 10, and AΔR reaches its maximum value. Because the data for Co in Fig. 6 continued to rise with increasing $t_{Co}$ to the largest value of $t_{Co}$ used, they were taken as setting only a lower bound ~ 40 nm on $l_{sf}^{Co}$ [47]. We note for later use that we find this $l_{sf}^{Co}$ to be less certain than $l_{sf}^{CoFe}$ or $l_{sf}^{Py}$. The slower growth of the Co data in Fig. 6 with increasing $t_{Co}$ makes it harder to be sure just where saturation occurs. For example, if data taking had stopped at $t_{Co}$ = 30 nm, the data could have been interpreted as saturating after $t_{Co}$ = 20 nm. In addition, for the thickest $t_{Co}$, the pinning field becomes close to the reversing field for the free Co layer with the same $t_{Co}$, so achieving a true AP state is less sure than for thinner $t_{Co}$.

### IIB2.b2. Determining $l_{sf}^{N}$ and $\delta_{N1/N2}$.

The basic sample geometry used to determine $l_{sf}^{N}$ and $\delta_{N1/N2}$ with EBSVs is the same as that used to determine $l_{sf}^{F}$, except that the common ferromagnetic layer thickness $t_F$ is now held fixed, and a new entity, X, is inserted into the middle of the central Cu layer. To determine $l_{sf}^{N}$, X is a single N-metal layer, X = N, as shown in Fig. 7a [48]. In Fig. 7a, I designates the interfaces, which are treated for convenience as additional thin layers. To determine $\delta_{N1/N2}$, X is a multilayer, X = [N1(3)/N2(3)]$_N$, where the common thickness, $t_{N1} = t_{N2}$ = 3 nm, is chosen to be larger than typical interface thicknesses (~0.6-1.0 nm) [49], so that N1 and N2 represent mostly 'bulk' material, yet small enough so that the spin-flipping due to finite $l_{sf}^{N1}$ and $l_{sf}^{N2}$ is generally small compared to that due to $\delta_{N1/N2}$. For most interfaces studied so far, N2 = Cu, simplifying the analysis. The middle of the sample then looks like Fig. 8a [48].

### IIB2.b2.a. X = N

Inserting X = N has two effects upon the EBSV, first adding thickness $t_N$ of N, and second adding two N/Cu interfaces. If, first, we neglect the two interfaces, VF theory can be approximated by [48]:

$$A\Delta R \: \alpha \: \exp[-t_N / l_{sf}^{N}]/(AR_o + AR_N). \tag{11}$$

Here $AR_o$ is the contribution to the denominator from the EBSV without the insert, $AR_N$ is the specific resistance increase due to the insert N, and the constant of proportionality depends upon the bulk and interfacial spin asymmetry parameters for Py. When $t_N \ll l_{sf}^{N}$, $AR_N$ is just $\rho_N t_N$ (still neglecting the interfaces), which increases linearly with $t_N$. When $t_N \gg l_{sf}^{N}$, $AR_N = \rho_N l_{sf}^{N}$, a constant. Thus, strictly, a simple exponential decay occurs only for $t_N \gg l_{sf}^{N}$.

Including the two N/Cu interfaces complicates Eq. (11), as described in [48]. Fig. 7b shows the resulting variation of log ($A\Delta R$) vs $t_X$ for $l_{sf}^{N}$ = 10 nm and neglecting any spin-flip scattering at the N/Cu interfaces. For a detailed fit to experimental data, such interfacial spin-flipping must be included, but it doesn't change the qualitative form of the curve. The initial rapid decay of log ($A\Delta R$) in Fig. 7b is due to the formation of the two N/Cu interfaces, which contribute to the term $AR_N$ in the denominator of Eq. (11) (and can also add interfacial spin-flipping). The slower, longer range decay comes mostly from the exponential term in Eq. (11) after the interfaces have completely formed. When $t_N > l_{sf}^{N}$, the slope of the full curve approximates that of Eq. (11) with a constant denominator—the dashed curve in Fig. 7b.

Fig. 9 [48] shows examples of log ($A\Delta R$) vs $t_N$ for several different N. The residual resistivities, $\rho_o$, determined from separately sputtered thin films of the metals, are given in Table II. In two cases, there is little or no interfacial contribution: (a) the dilute alloy N = Cu(6%Pt), which should have no real interface with Cu; and (b) N = Ag, where both the Cu/Ag interface specific resistance and interfacial spin-flipping are small [48, 49]. For Cu(Pt)/Cu, log ($A\Delta R$) vs $t_N$ is close to a single exponential, dominated by the contribution from $l_{sf}^{N}$. For Cu/Ag, $l_{sf}^{N}$ is long enough that the variation of log ($A\Delta R$) is dominated by the $AR_N$ in the denominator of Eq. 11. In contrast, V, Nb, and W, all have relatively large interface specific resistances [48], but small to large interfacial spin-flipping (see Table IV). In these cases, the additional resistances (and spin-flipping) produced as the interfaces form, dominate the initial decrease of $A\Delta R$ as $t_N$ increases from $t_N$ = 0, leading to a rapid falloff of log ($A\Delta R$) with increasing $t_N$. Only after the interfaces are fully formed, should the rate of falloff decrease to close to that due to $l_{sf}^{N}$ alone. Data such as those for V, Nb, and W in Fig. 9 can be analyzed for $l_{sf}^{N}$ either by fitting the data for large $t_N$ to the single exponential exp(-$t_N$/$l_{sf}^{N}$) of Eq. 11, or by making a complete fit with VF theory taking account of the interface specific resistance and interfacial spin-flipping. To determine these additional parameters requires a simultaneous fit to data with inserts of [N/Cu]$_N$ interfaces, as we discuss next. In the range of thicknesses initially studied, the decreases of the data for V and Nb beyond the 'knees' in Fig. 9 are so slow (i.e., comparable to what is expected just from the additional term $AR_N$ in the denominator of Eq. (11)), that only lower bounds on $l_{sf}^{N}$ could be derived. Extensions to thicker layers provided the values of $l_{sf}^{V}$ [50] and $l_{sf}^{Nb}$ [51] listed in Table II. Finally, the spin-flipping in FeMn in Fig. 9 is so strong that it could not be distinguished from just interfacial spin-flipping.

**IIB2.b2.b.  X = [N1/N2]$_N$**

To determine AR$_{N1/N2}$ and $\delta_{N1/N2}$ one uses a multilayer insert of the form X = [N1(3)/N2(3)]$_N$, with fixed layer thicknesses of 3 nm for both N1 and N2 [Fig. 8a]. VF theory can be approximated by:

$$A\Delta R \propto \exp[-2\mathcal{N}\delta_{N1/N2} - \mathcal{N}(3/l_{sf}^{N1}) - \mathcal{N}(3/l_{sf}^{N2})]/(AR_o + AR_X), \qquad (12)$$

where AR$_X$ is the contribution of the insert X, and exponential decay is due to spin-flipping at the interfaces and also within the N1 and N2 layers. In Fig. 8b we compare A$\Delta$R for Eq. (12)—dashed curve, with a more complete fit to the VF equations—solid curve, where for simplicity in both cases we took $l_{sf}^{N1} = l_{sf}^{N2} = \infty$. Eq. (12) then approximates the slope of the solid curve for $\mathcal{N}$ greater than 'a few'. Fig. 10 [48] shows plots of log (A$\Delta$R) vs $\mathcal{N}$ for multilayer inserts of X = [Ag/Cu]$_N$, [V/Cu]$_N$, [Nb/Cu]$_N$, and [W/Cu]$_N$. Values of 2AR$_{N1/N2}$ and $\delta_{N1/N2}$ for these and other interfaces are given in Table IV. The procedure used to determine $\delta_{N1/N2}$ with EBSVs is the same as that used to determine $l_{sf}^{N}$, and its use for $l_{sf}^{N}$ has been validated as discussed at the end of Section IIB2.a. It, thus, seems likely to be valid for $\delta_{N1/N2}$. However, there are as yet no independent measurements or calculations of $\delta_{N1/N2}$. Thus the fundamental significance of the values listed is not yet clear. We don't know if different techniques will give similar values, or if the values given are sensitive to interfacial structure and/or intermixing.

Strictly, this procedure works only for two non-magnetic (N) metals, since inserting an F-metal into the middle of the EBSV fundamentally changes its magnetic structure. It might be possible to keep the direction of magnetization of such a middle layer fixed, and thereby simplify the problem enough to extract spin-flip information at an F/N interface, but this procedure has not yet been implemented. For the moment, there is no established technique for reliably measuring $\delta_{F/N}$ for F/N interfaces. The first inference of a non-zero $\delta_{F/N}$ ($\delta_{Co/Cu} \sim 0.25$ at 4.2K) was made in [52] to rationalize, within the VF theory, the difference between data for 'interleaved' and 'separated' Co/Cu multilayers as described in detail in Appendix C This same value was later shown to help explain both the CPP-MR of Co/Cu EBSVs [16] and effects of adding internal interfaces (laminating) on CPP-MR [53]. We note in passing that the difference in AR data for interleaved and separated samples of Co/Ag in Fig. 16 of Ref. [16] suggests that a similar analysis would give a roughly similar value for $\delta_{Co/Ag}$. While, together, these three studies strongly suggest a non-zero $\delta_{Co/Cu}$, they are not quite definitive, because they assume a long $l_{sf}^{Co}$ (which, as we note in IIB2.b1 and IIB2.c1, is probable, but not absolutely sure), and they infer a non-zero $\delta_{Co/Cu}$ to achieve another goal, not from measurements designed explicitly to detect it. Three additional studies inferring non-zero values of $\delta_{F/N}$ have recently been published. Strong spin-flipping ($\delta_{Py/Cu} \sim 0.95$) at Py/Cu interfaces at 293K was inferred from failure of an LNL signal to be as large as expected [54]. We worry that the model and parameters are not well enough established to reach this conclusion. Values of $\delta_{F/N} \approx 0.3$ at 293K for F/N = Co/Ru and Py/Cu were derived indirectly from fits to CPP-MR data in nanopillars [55] {Note: first author A.M. of [55] informs us that the value in Table I of [55] of $\delta_{F/N}$ = 0.25 for Co$_{90}$Fe1$_{10}$ was simply taken equal to that for Co/Cu (we agree with this assumption), and that the value of $\delta_{F/N}$ = 0.33 listed for Ni$_{50}$Fe$_{50}$ was actually derived for the Py data of [15]}. A value of $\delta \sim 0.5$ for Co$_{50}$Fe$_{50}$ at 4K and 300K was derived from fits to A$\Delta$R and AR data for nanopillars with laminated (internal interfaces) Co/Cu layers [56]. These latter two studies used bulk and interface parameters from CPP-MR/S measurements at 4.2K, taking them to be temperature independent. While a weak temperature dependence is plausible [57], complete temperature independence seems less likely, and it isn't clear that the CPP-MR/S parameters will all be quantitatively applicable to nanopillars with different layer residual resistivities and microstructures. Also, in the laminated study, the Cu layers were so thin (0.3 nm) [56] that it is not clear that the interfaces were fully formed and independent [53]. Taken together, these more recent results modestly strengthen the case for a non-zero $\delta_{F/N}$. Finally, a recent paper presents a potential way to use LNL measurements to derive information about $\delta_{F/N}$ as a function of temperature [58]. Combined with the data for $\delta_{N1/N2}$ in Table IV, these results suggest that some spin flipping at F/N interfaces is likely. However, because the derivations of non-zero $\delta_{F/N}$ are indirect, and most depend strongly upon assumptions about models and parameters that are not clearcut, we view them (with the possible exception of Co/Cu) as highly uncertain. We, thus, describe these studies, but do not collect the inferred values of $\delta_{F/N}$ into a separate table. Non-zero values of $\delta_{F/N}$ would also require a source. One such source is spin-orbit scattering. A crude spin-orbit argument in Ref. [16] produced $\delta_{Co/Cu} \sim 0.2$ for a 50%-50% Co/Cu interface alloy, a value comparable to the inferred $\delta_{Co/Cu}$ = 0.25. Another such source is moment non-collinearity at an F/N interface. Such non-collinearity at Py/Cu interfaces has been proposed [59], but apparently not at Co/Cu ones [59]. We conclude that spin-flipping at [F/N] interfaces requires more study, both experimental and theoretical.

### IIB2.c. CPP-NW/ML

The second geometry used to find $l_{sf}^F$ and $l_{sf}^N$ from the CPP-MR involves nanowires. This geometry has the advantage that measurements can be extended to room temperature.

**IIB2.c1.** For $l_{sf}^F$, an inverted form of Eq. 7a can be generalized using the VF equations to include $l_{sf}^F$. In the limits $t_F \gg l_{sf}^F$; $t_N \ll l_{sf}^N$, one obtains [25, 60]:

$$R_P/\Delta R = [(1-\beta_F^2)t_F]/(2p\beta_F^2 l_{sf}^F) \tag{13}$$

Here, because it was not clear how close the nanowire multilayers approached a true AP configuration, the parameter p was introduced as the fraction of the AP configuration between adjacent layers (p = 1 is AP). Fig. 11 [43] shows a plot of $R_P/\Delta R$ vs $t_{Co}$. Later analysis in [61], assuming p = 0.85 and $\beta_{Co}$ = 0.36, gave $l_{sf}^{Co}$ = 59 nm at 77K and 39 nm at 295K as listed in Table III. These values are likely quite uncertain—e.g., choosing $\beta_{Co}$ = 0.46 [11, 62], would give $l_{sf}^{Co}$ = 33 nm at 77K and 22 nm at 295K

For Py, to ensure p close to 1, the multilayers were made with alternating thin (10nm) and thick (100 or 500 nm) Cu layers. The dipolar interaction then coupled antiferromagnetically the Py layers separated by only 10 nm, but these coupled Py pairs were uncoupled from each other. The resulting $l_{sf}^{Py}$ = 4.3 nm at 77K is listed in Table III [25]

**IIB2.c2.** The equation needed to derive $l_{sf}^N$ with nanowires is more complex, necessitating a detailed numerical fit to the data. The value of $l_{sf}^{Cu}$ listed in Table II for this method is shorter than those found by other methods, probably because electrodeposition gives 'dilute alloys' of Co and Cu (alloying probably worse in the Co than in the Cu) rather than pure metals [63, 64], and the additional scattering reduces $l_{sf}^{Cu}$.

### IIB2.d CPP-NP.

So far, there is only one example of nanopillars used to derive $l_{sf}^N$ [65], giving a RT value of $l_{sf}^{Cu}$ = 190 ± 20 nm. As shown in Table II, this value is shorter than most of those derived by other methods; the reason why is not yet clear.

### IIC Lateral (L) and non-local (NL) geometries.

The advantages of experiments with lateral (L) geometry are that they can be carried out at both room and low temperatures, and they can be used for long $l_{sf}^N$. There are, however, also disadvantages in some published studies if F/N interfaces are low resistance: (a) the current density is not uniform, and (b) the equations used to analyze the data are complex and there is disagreement about the form(s) to use. For these reasons, some published values from LNL studies look uncertain. As we explain below, the most reliable are Hanle studies and ones with high resistance (e.g., tunneling or very dirty metal) contacts.

The geometries used in L- and LNL experiments are shown in Fig. 12. The quantity measured is usually $\Delta R = R(AP) - R(P)$, the difference in resistance between states where the magnetizations of the F-layers F1 and F2 are AP or P to each other. The models used to analyze most published LNL data are 1D models that assume that uniform charge and spin currents flow through the sample. Refs. [54, 66] showed that this assumption is violated when the F/N interfaces are metallic (i.e., low or moderate resistance) and the length $L$ of the N-metal of interest is not $\gg W_N$, the width of the N-metal. For such interfaces, the current injected from F1 into N is non-uniform because the lower resistivity N-metal partly shorts out the higher resistivity F-strip [54, 66, 67]. Only when $L \gg W_N$ is the resulting initially non-uniform spin-current able to become more nearly uniform by the time it reaches the other cross-strip, F2. Problems with non-uniform currents are exacerbated by application of a magnetic field (see, e.g., [68, 69]). In addition, several studies with metallic interfaces used an equation (e.g., Eq. 15 below) that doesn't properly take account of the interface resistances as in Eq. 16a. We'll see that use of Eq. 15 could yield too small a value of $l_{sf}^N$. Because the techniques used in most LNL studies with metallic F/N interfaces involve the possibility of contamination of those interfaces during preparation, it is important for authors to independently measure and specify their F/N interface resistances to show if they are less than, comparable to, or greater than those of the F- and N-metals (see discussion associated with Eq. 16 below).

Since this is not a theoretical review (See, e.g., [28]), we do not go into details of models and analyses, but focus only on the different equations that are most relevant to an experimentalist.

We begin with the lateral (L) geometries shown in Fig. 12.  Fig. 12a shows a lateral thin-film magnetoresistance geometry (LMR) used for Standard (a-#1) or LNL (a-#2) measurements.  Because the a-#1 geometry gives difficulties in eliminating unwanted anisotropic MR and Hall effect contributions [70], most studies use the LNL geometry shown as a-#2 or variants thereof.  Figs. 12b, 12c, and 12d show three such variants.  We label Fig. 12b with its cross as LNL/C and Fig. 12c with extra strips as LNL/+.  Finally, Fig. 12d shows what is called a Lateral Three-Terminal device (LNL/TT).  If the contacts between the F and N layers are tunneling contacts, we label the sample as LNL/T.  A lateral measurement involving a spin-dependent Hall Effect (L/SDHE) will also be noted.

### IIC1. LNL with the Hanle Effect (LNL/H).

The first geometry used to determine $l_{sf}^N$ was the LNL geometry of Fig. 12a #2, combined with the Hanle effect— LNL/H [71]. $l_{sf}^{Al}$ was measured on a rolled and then annealed Al foil with residual resistance ratio RRR = R(RT)/R(4.2K) = 1100, much higher than that of any other sample covered in this review.  In the Hanle effect, a magnetic field, $B_\perp$, applied perpendicular to the sample plane, causes the moments of electrons to precess as the electrons move from F1 to F2.  Because diffusive transport gives a broad distribution of the number of electrons arriving at F2 as a function of time, and the longer the time, the more likely for the moment to have flipped, the precession causes the voltage at F2 (whose moment is aligned parallel to that of F1) to decrease with increasing $B_\perp$ (more precisely with increasing Larmour frequency, $\omega_\perp = (g\mu_B B_\perp)/\hbar$, where $g \sim 2$ is the electron g-factor, $\mu_B$ is the Bohr magneton, and $\hbar$ is Planck's constant divided by $2\pi$).  The detailed equation, which is fit numerically, is given in [72].  Whereas the first Hanle measurements were made with metallic F/N contacts, a more recent study of Al was made with tunneling contacts.  In that case, the values of $l_{sf}^{Al}$ found from LNL/T and LNL/H were closely the same [73].

### IIC2. LNL with Three Terminal Geometry (LNL/TT)

The second geometry tried involved three-terminals (TT)—12d.  It was used for Au [74] and Nb [75].  The current flow in this geometry cannot be uniform, and non-uniform currents can lead to unwanted magnetoresistive effects [68, 69].  As shown in Table II, the values of $l_{sf}^{Au} \sim 1.5$ μm and $l_{sf}^{Nb} \sim 0.8$ μm inferred from this geometry are both an order of magnitude larger than those for samples of comparable purity found with other techniques.  For additional issues see [76].

### IIC3. LNL with a Cross Geometry (LNL/C)

The next studies used a non-local geometry involving a cross (Fig. 12b), LNL/C, for both Cu [70, 77] and Al [77].  Assuming low resistance, metallic interfaces, the data were analyzed using Eq. 14.

$$\text{For General L:} \quad \Delta R = [\beta_F^2 R_N e^{-L/2l_{sf}^N}]/(M+1)[M\sinh(L/2l_{sf}^N) + \cosh(L/2l_{sf}^N)] \tag{14}$$

where $M = (A_N R_N / A_F R_F)(1-\beta_F^2) = (\rho_N l_{sf}^N / \rho_F l_{sf}^F)(1-\beta_F^2)$ and $\beta_F$ was defined above.  For Py and Cu, M ~ [(10)/(0.7)](0.5-0.7) ~ 10.  With such a large M, the sinh dominates, and in the experimental limit $L \geq l_{sf}^N$, Eq. 14 becomes similar to our preferred Eq. 16a, below, but with a few differences due to the different geometry (cross vs standard non-local).  As in Eq. 16a, the size of $\Delta R$ is determined not by $\beta_F^2$ alone, but a product, here $\approx [\beta_F (\rho_F l_{sf}^F)/\{(1-\beta_F^2)(\rho_N l_{sf}^N)\}]^2$, that can be $\ll \beta_F^2$

These pioneering studies of e-beam fabricated samples with metallic interfaces were criticized [67] for: (a) non-uniform current injection from Py into the N-metal; (b) neglect of interface resistances, which [67] claimed should dominate; and (c) possible unwanted contributions from anisotropic magnetoresistance.  The correctness and significance of these arguments was strongly disputed [78].  As we have noted above, (a) represents a potential problem for all LNL measurements with metallic interfaces.  But in the present experiments, its effect was mitigated by using samples with $L \gg W$ and mostly $L \geq l_{sf}^N$.  The resulting values of $l_{sf}^N$ are competitive (see Table II and Figs. 14 and 16).  But, as noted in [77], non-uniform current injection can still affect the inferred $\beta_F$ by a factor ~ 2-3.

### IIC4. LNL with Metallic Interfaces (LNL/M)

Some subsequent studies involved a non-local geometry with metallic interfaces and without a cross.  As noted above, unless these have $L \gg W$, the 1D equations used for analysis may not be applicable because of a current uniformity problem.  For two [79, 80], there is also another issue, involving the equation used for analysis.  This equation was:

$$A\Delta R = [(P_1 P_2 R_N)](e^{-L/l_{sf}^N}). \tag{15}$$

Here $P_1$ and $P_2$ are injector and detector spin polarization values, and $R_N = \rho_N l_{sf}^N /A$, where A is the cross-sectional area of the N-stripe. From Eq. 15, one should obtain $l_{sf}^N$ from the slope of a straight line on a plot of log (A$\Delta$R) vs $L$.

More recently, Takahashi et al. [81, 82] generalized the 1D equations to explicitly include the F/N interface resistances $R_i = AR_{Fi/N}/A_J$. Here $AR_{Fi/N}$ is the AR for the $F_i$/N interface, and $A_J$ is the (junction) area of overlap of the $F_i$ and N layers in the LNL geometry (Fig.12a#2). They get:

(1) $R_i = R_1, R_2 \ll R_F \ll R_N$.   $\Delta R = [4 p_F^2/(1-p_F^2)^2][R_N(R_F/R_N)^2][e^{-L/l_{sf}^N}/(1-e^{-2L/l_{sf}^N})]$   (16a)

(2) $R_F \ll R_i \ll R_N$.   $\Delta R = [4 P_J^2/(1-P_J^2)^2][R_N(R_1 R_2/R_N^2)][e^{-L/l_{sf}^N}/(1-e^{-2L/l_{sf}^N})]$   (16b)

(3) $R_i \gg R_N \gg R_F$   $\Delta R = P_J^2 R_N e^{-L/l_{sf}^N}$   (16c)

Here $p_F$ is the polarization within the F-metal (equivalent to $\beta_F$ defined above), $P_J$ is the polarization of the F/N interface (equivalent to $\gamma_F$ defined above); $R_i = AR_{Fi/N}/A_J$, with i,J = 1,2; $R_F = \rho_F l_{sf}^F /A_F$; and $R_N = \rho_N l_{sf}^N /A_N$. To estimate $R_F$, $R_N$, and $R_i$ for clean metallic interfaces, we use values from Tables II and III and ref. [11]. At 4.2K, a Cu strip of W = 100 nm, t = 40 nm, combined with $\rho_{Cu} l_{sf}^{Cu} = 20$ f$\Omega$m$^2$, gives $R_N \approx 5$ $\Omega$. At 4.2K, a Py strip of W = 100 nm, t = 40 nm, and $\rho_{Py} l_{sf}^{Py} = 0.7$ f$\Omega$m$^2$ gives $R_F \approx 0.2$ $\Omega$. Eq. (16a) would then be appropriate for a metallic Py/Cu interface with $AR_{Py/Cu} = 0.5$ f$\Omega$m$^2$ [11] and A = (100 nm)$^2$, giving $R_i = 0.05$ $\Omega$. However, as noted above, contamination of the F/N interface during sample preparation could increase $R_i$. To justify using Eq. 16c, $R_i$ would have to be more than 100 times larger than our estimate.

Note that only Eq. 16c for high resistance (e.g., tunneling or very dirty metallic) interfaces contains just a single exponential. Both Eqs 16a for low resistance metallic interfaces, and 16b for intermediate resistance interfaces, are more complex, reducing to a single exponential only when $L \gg l_{sf}^N$. Fig. 13 [73, 82-84] should make the difference clear. The upper two sets of data, for tunneling interfaces, are consistent with the single exponentials predicted by Eq. 16c. While the data for the lower two sets, for metallic interfaces, could be approximately fit by single exponentials as per Eq. 15, fits to Eq. 16a (filled squares and solid curve: $l_{sf}^N \sim 920$ nm at 4.2K; open squares and dashed curve, $l_{sf}^N \sim 700$ nm at 295K) show more complex behavior, where the data should vary as a single exponential only when $L \geq 2 l_{sf}^N$. If Eq. 16a is correct, then analyzing data with clean metallic interfaces assuming just a single exponential (Eq. 15) can give an incorrect $l_{sf}^N$ (too short if $L$ does not extend far enough), as well as incorrect values for the polarization $p_F$. We emphasize that to derive $p_F$ from Eq. 16a requires knowledge of both $R_N$ and $R_F$, and to derive $l_{sf}^F$ requires knowledge of both $R_N$ and $p_F$. An inferred too-short $l_{sf}^N$ could also mislead about whether the data are in the correct regime for the single exponential limit of Eq. 16a.

### IIC5. LNL with Multiple Cross-Strips (LNL/+)

We use this acronym for LNL studies (such as [84-87]) that involve additional N2 (Cu or Au) or F3 (Py) strips crossing the main N(Cu) strip as in Fig. 12c. Multiple F-strips have the potential advantage that different values of $L$ are all associated with the same N-strip, instead of with different N-strips that might have different impurity contents. However, Kimura et al. [84, 85] reported that, for low resistance metallic interfaces, the presence of additional strips affects the analysis, reducing the magnitude of the signal by allowing 'spin-accumulation to leak out through these additional leads'. Assuming a simplified 1D analysis, they derived an equation relating the resistance $R_{SM} = R_M l_{sf}^M /A_M$ of an extra strip to the measured $\Delta$R. From this analysis they derived values of $l_{sf}^M$ listed in Tables II and III [84]. Because their analysis does not explicitly include the interface resistances as in Eq. 16, we are not sure how reliable it is. Their results are similar, but not identical, to those found in other ways. In contrast, Godfrey & Johnson [86], who studied $l_{sf}^{Ag}$ using sets of four separate, but different width Py-strips on a single Ag strip, claimed that their tests showed no direct effects of extra strips outside of their uncertainties. But their value of $l_{sf}^{Ag}$, may have been affected by their use of a single exponential fit. Also, effects of extra leads might have been reduced by interface resistances ($AR_{Py/Ag} = 2.4$ f$\Omega$m$^2$) about five times larger than our estimate above. For reasons similar to why we worry about Eq. 15, we worry also about the equation that they used to estimate $l_{sf}^{Py}$, which gave an outlying value in Table III for samples with comparable values of $\rho_{Py}$. Lastly, Ku et al. [87] measured $l_{sf}^{Au}$ using multiple Py strips across the Au. They found no effect of the multiple strips, but their reported Py/Au interface resistance was unusually large ($AR_{Py/Au} = 110$ f$\Omega$m$^2$).

**IIC6. LNL with Tunneling Interfaces (LNL/T)**

Finally, the last acronym, LNL/T designates LNL measurements (so far only of $l_{sf}^{Al}$) involving tunneling F/N interfaces made by oxidizing the surface of the Al before the F-layers were deposited. Ref. [73] compared the results of LNL/T measurements with those of LNL/H measurements on samples with $L >> W_N$. The two techniques gave very similar results. The combination of tunneling interfaces, $L >> W_N$, and Hanle measurements, looks to be especially reliable. By comparing LNL/T data for different thicknesses of Al films, Ref. [88], measured $l_{sf}^{Al}$ and deduced that spin-relaxation at 4K is weaker at the film surface than in the bulk. Ref. [89] gives values of $l_{sf}^{Al}$ and shows SHE data.

**IIC7. LNL with Spin-Dependent Hall Effect (LNL/SDHE).** An $l_{sf}^{Al} \sim 5$ μm at 2K was inferred from spin-dependent Hall Effect measurements [90].

**IID1. $l_{so}$ and $l_{sf}^N$ from Weak-localization (WL).**

To obtain a large Weak-Localization (WL) signal, magnetoresistance measurements for WL analysis are made at T ≤ 40K on evaporated or sputtered thin films, sometimes quench-condensed to increase the residual resistivity. WL measurements can be in 'quasi-1D', 'quasi-2D', or 3D regimes, depending upon whether the phase coherence length, $\lambda_\phi$, is larger than both $W$ and $t$ (1D), just $t$ (2D), or none of $L$, $W$, and $t$ (3D). We include in this review, data in the 2D and 1D regimes of 'nominally pure metals'. In both 2D and 1D, spin-orbit scattering changes the sign of the WL contribution to the MR. The spin-orbit length, $l_{so}$, can be determined from a WL equation if the sample wires are in the diffusive regime and far from the metal-insulator transition [91]. In practice, if $l_{so} >> \lambda_\phi$, or $l_{so} << \lambda_\phi$ the data just determine $\lambda_\phi$. So one must choose a measuring temperature so that the rapidly varying $\lambda_\phi$ is comparable to the temperature independent $l_{so}$. If spin-orbit scattering is dominant, $l_{so} = l_{sf}^N$. We list WL values of $l_{so} = l_{sf}^N$ for Ag, Au, Cu, Al, and Mg in Table II. Values of $l_{so}$ for Mg-based alloys can be found in [92, 93].

**III. Data Tables and Comments.**

Section III contains four Tables. For each, we briefly discuss the data it contains and the limitations thereon. Table I compares, with independent calculations, experimental values of $l_{sf}^N$ for alloys in which the scattering is dominated by known concentrations of known impurities. By these comparisons, we test quantitatively both the VF theory of CPP-MR, and two CPP-MR techniques for measuring $l_{sf}^N$, one used again in Table II. Table II lists experimental values of $l_{sf}^N$ for nominally pure metals at temperatures T ranging from 4.2K to 293K and obtained by several different techniques. Comparing techniques is important, as the values of $l_{sf}^N$ are generally not intrinsic to a given metal. To see if $l_{sf}^N$ is proportional to the inverse transport mean-free-path, $\lambda_t$, we plot in the 6th column of Table II the product $\rho\, l_{sf}^N$. Table III lists experimental values of $l_{sf}^F$ for a series of nominally pure F-metals and F-based alloys, along with the product $\rho_o l_{sf}^F$. Table IV lists experimental values for $\delta_{N1/N2}$.

The techniques used in these Tables are listed by acronyms:

| | |
|---|---|
| CPP-S/ML. | CPP-MR using Superconducting cross-strips and multilayers. |
| CPP-S/SV | CPP-MR using Superconducting cross-strips and exchange-biased spin-valves. |
| CPP-NW | CPP-MR using electrodeposited nanowire multilayers. |
| CPP-NP | CPP-MR using electron-beam lithography produced nanopillar trilayers. |
| WL | Weak Localization. |
| LNL/M | Lateral Non-local MR with metallic contacts and no other special conditions.. |
| LNL/H | Lateral Non-local MR using the Hanle effect. |
| LNL/C | Lateral Non-local MR using a cross-geometry for the N-metal. |
| LNL/+ | Lateral Non-local MR with an extra strip or strips across the N-metal. |
| LNL/T | Lateral Non-local MR using tunneling contacts. |
| LNL/TT | Lateral Non-local MR using three terminals. |
| LNL/SDHE | Lateral Non-local Spin-Dependent Hall Effect |

**IIIA. Table I. Spin-Diffusion Lengths, $l_{sf}^N$, in Non-magnetic Alloys**.

Table I contains values of $l_{sf}^N$ at 4.2K for alloys with known concentrations of specific impurities. The values were found by two different techniques--CPP-S/ML (Section II.B.2a.1) and CPP-S/SV (Section II.B.2b)--the results of which agree in the two cases of overlap. As the scattering in each alloy is dominated by a single source, each $l_{sf}^N$ should be intrinsic to that alloy. For impurities without a local magnetic moment, $l_{sf}^N$ should be dominated by spin-orbit scattering and thus calculable from CESR-derived spin-orbit cross-section for that impurity in the given host [26] plus the constant $\rho_b l_b$ for that host, as described by Eqs. 3-5. All measured and calculated values are in good agreement. For impurities with a local moment, more complex spin-spin calculations are needed [27]. Here, too, the measured and calculated [27] (labeled by * in Table I) values agree. The agreements between values of $l_{sf}^N$ found by two different techniques, and between measured and calculated values, together suggest that the VF theory used to analyze the data is appropriate, and that both experimental techniques correctly measure $l_{sf}^N$. In contrast, the fact that larger $l_{sf}^N$ s for AgSn and CuGe are associated with larger residual resistivities, $\rho_o$, while smaller $l_{sf}^N$ s for AgPt and CuPt, are associated with smaller $\rho_o$s, makes it hard to see how 'mean-free-path' effects alone (Appendix C) could explain these values or the data in Figs. 3, 4, 9. Explaining these data seems to require spin-flipping.

No uncertainties are given for most of the values of $l_{sf}^N$ measured with multilayers or for the calculations. For the measurements, fluctuations in the data of Figs. 3 and 4 make it difficult to reliably estimate uncertainties. For the calculations, there is some uncertainty in the constant $\rho_b l_b$ used to determine $\lambda_t$ (see Section II.A and Appendix B).

| Alloy | T(K) | Technique | $l_{sf}^N$ (nm)(exp) | $l_{sf}^N$ (nm)(CESR) | $\rho_o$(nΩm) | Ref. |
|---|---|---|---|---|---|---|
| Ag(4%Sn) | 4.2 | CPP-S/ML | ≥ 26 | | 200±20 | [12] |
| Ag(6%Pt) | 4.2 | CPP-S/ML | ≈ 10 | ≈ 7 | 110±20 | [12] |
| Ag(6%Mn) | 4.2 | CPP-S/ML | ≈ 11 | ≈ 12* | 110±25 | [12] |
| Ag(9%Mn) | 4.2 | CPP-S/ML | ≈ 7 | ≈ 9 * | 155±20 | [12] |
| Cu(4%Ge) | 4.2 | CPP-S/ML | ≥ 50 | ≈ 50 | 182±20 | [45] |
| Cu(6%Pt) | 4.2 | CPP-S/ML | ≈ 8 | ≈ 7 | 130±10 | [12] |
| Cu(6%Pt) | 4.2 | CPP-S/SV | 11±3 | ≈ 7 | 160±30 | [48] |
| Cu(7%Mn) | 4.2 | CPP-S/ML | ≈ 2.8 | 3±1.5* | 270±30 | [12] |
| Cu(6.9%Ni) | 4.2 | CPP-S/ML | 23 | 22.4 | 110 | [94] |
| Cu(10%Ni) | 4.2 | CPP-S/ML | 14 | 14.7 | 175 | [94] |
| Cu(14%Ni) | 4.2 | CPP-S/ML | 10 | 11.9 | 191 | [94] |
| Cu(22.7%Ni) | 4.2 | CPP-S/ML | 7.5 | 6.9 | 355 | [94] |
| Cu(22.7%Ni) | 4.2 | CPP-S/SV | 8.2 ± 0.6 | 7.4 | 310±20 | [95] |

* Values calculated for spin-spin scattering in [27].

**IIIB. Table II. Spin-Diffusion Lengths, $l_{sf}^N$, in nominally 'pure' non-magnetic metals.**

Table II contains values of $l_{sf}^N$ for nominally 'pure' non-magnetic metals found in several different ways. For each metal, the values are listed in chronological order. The values at room temperature should be intrinsic if scattering by phonons dominates the resistivity. To show the reader where this intrinsic value would be expected, we list in the captions to Figs. 14-16 the values of $(\rho_N)^{-1}$ at 293K for high purity N = Cu (Fig. 14), Ag (Fig. 15), and Al (Fig. 16). If, however, a substantial fraction of that resistivity is due to defects, $l_{sf}^N$ won't be intrinsic. In contrast, the values at low temperatures (4.2K, 10K) cannot be intrinsic, since the lattice defects or impurities that dominate the scattering are unknown. To test the hope that samples with similar residual resistivities might be dominated by similar impurities, giving similar values of $l_{sf}^N$ that scale roughly inversely with the resistivity, $\rho$, the next to last column of Table II contains values of $\rho \, l_{sf}^N$. As explicit examples of such tests, Figs. 14 (for Cu), 15 (for Ag), and 16 (for Al) show plots of $l_{sf}^N$ vs $1/\rho$, including several individual samples where only ranges are given in Table II. Correlations in both figures are by no means perfect.

In general, comparing values in Table II for individual metals at either 4.2K or 293K shows substantial variations. These variations are due partly to variations in sample resistivity (see column $\rho \, l_{sf}^N$), but partly to experimental or analysis problems. As noted in section II.C, values of $l_{sf}^N$ from some LNL studies with metallic contacts are uncertain, due to use of inappropriate equations, and/or non-uniform current flow, and/or use of extra cross-strips.  As an example of effects of using different equations with the same data, compare 293K value of $l_{sf}^{Cu}$ = 500 nm derived by Kimura et al. [84] using a different equation with $l_{sf}^{Cu}$ = 700 nm derived by Takahashi et al. [82] using Eq. 16a (see Fig. 13).

| Metal | T(K) | Technique | $l_{sf}^N$ (nm) | $\rho_o; \rho(T)$(n$\Omega$m) | $\rho l_{sf}^N$ (f$\Omega$m$^2$) | Ref. |
|---|---|---|---|---|---|---|
| Au | 4.5 | WL | 10.5 | 665# | 7 | [96] |
| Au | ≤70 | LNL/TT | 1500±400 | | | [74] |
| Au | ≤ 4 | WL | 58.5 | 33# | 2 | [97] |
| Au | ≤ 4 | WL | 85 | 25# | 2 | [91] |
| Au | 4.2 | CPP-S/SV | $35^{+65}_{-10}$ | 19 ± 6 | $0.7^{+1.8}_{-0.4}$ | [98] |
| Au | 10 | LNL/M | 63±15 | | | [79] |
| Au | 293 | LNL/+ | 60 | 52 | 3 | [84] |
| Au | 15 | LNL/+ | 168 | 40 | 7 | [87] |
| Cu | 4.5 | WL | 39 | 720# | 28 | [96] |
| Cu | 77 | CPP-NW | 140±15 | 31 | 4 | [43] |
| Cu | 300 | CPP-NW | 36±14 | 20-65 | 0.4 -3 | [64] |
| Cu | 4.2 | LNL/C | 1000±200 | 14 | 14 | [70, 77] |
| Cu | 293 | LNL/C | 350±50 | 29 | 10 | [70, 77] |
| Cu | ≤ 4 | WL | 520 | 35# | 18 | [97] |
| Cu | ≤ 4 | WL | 330-670 | 17-48(C)#. | 11-16 | [91] |
| Cu | 293 | CPP-NP | 170±40 | | | [65] |
| Cu | 293 | LNL/M;LNL/+ | 500 | 21 | 11 | [84] |
| Cu | 293 | LNL/M | 700 | 21 | 15 | [82] |
| Cu | 4.2 | LNL/M | 920 | 34 | 31 | [82] |
| Cu | 4.2 | LNL/H | 546 | 34 | 19 | [58] |
| Cu | 10 | LNL/M | 200±20 | 13.6 | 3 | [80] |
| Cu | 300 | LNL/M | ≥ 110 | 34 | ≥4 | [80] |
| Al | ~ 4K | WL | 300-570 | 23-68# | 11-20 | [24, 99] |
| Al | 4.3 | LNL/H | 450,000 | ~0.024 | 11 | [3, 71] |
| Al | 37 | LNL/H | 170,000 | ~0.024 | 4 | [3, 71] |
| Al | ~ 4K | WL | 450-560 | 22-36# | 12-16 | [100] |
| Al | 4.2 | LNL/T | 650 | 59 | 38 | [73] |
| Al | 293 | LNL/T | 350 | 91 | 32 | [73] |
| Al | 4.2 | LNL/C | 1200 | 13 | 16 | [77] |
| Al | 293 | LNL/C | 600 | 32 | 19 | [77] |
| Al | <100K | L/SDHE | ~5000 | 7.8 | 39 | [90] |
| Al | 2K | LNL/T | 400±50 | | | [101] |
| Al | 293 | LNL/T | 350±50 | | | [101] |
| Al | 4K | LNL/T | 660 | 20# | 13 | [88] |
| Al | 293 | LNL/T | 330 | | | [88] |
| Al | 4.2 | LNL/T | 455±15 | 95# | 43 | [89] |
| Al | 4.2 | LNL/T | 705±30 | 59# | 42 | [89] |
| Ag | 4.5 | WL | 26-33 | 440-830# | 26-33 | [96] |
| Ag | 4.2 | CPP-S/SV | > 40 | 7±2 | >0.3 | [48] |
| Ag | ≤ 4 | WL | 750** | 30# | 23 | [97] |
| Ag | ≤ 4 | WL | 350-1000 | 21-55 (C)# | 19-21 | [91] |
| Ag | 79 | LNL/+ | 132-195 | 35-40 | 5-7 | [86] |
| Ag | 298 | LNL/+ | 132-152 | 49-55 | 7 | [86] |
| Cr | 4.2 | CPP-S/SV | ~ 4.5 | 180±20 | ~0.8 | [102] |
| Mg | ≤4 | WL | 80-220 | 860-5500# | 189-440 | [103] |
| V | 4.2 | CPP-S/SV | >40 | 105±20 | >4 | [48] |
| V | 4.2 | CPP-S/SV | 46±5 | 105±20 | 5 | [50] |

| Nb | 12K | LNL/TT | 780±160 | ~50 | 39 | [75] |
| Nb | 4.2 | CPP-S/SV | $25^{+\infty}_{-5}$ | 78±15 | 2 | [48] |
| Nb | 4.2 | CPP-S/SV | 48 ± 3 | 60±10 | 3 | [51] |
| Pd | 4.2 | CPP-S/SV | $25^{+10}_{-5}$ | 40 | 1 | [104] |
| Ru | 4.2 | CPP-S/SV | ~ 14 | 95 | 1.3 | [105] |
| Pt | 4.2 | CPP-S/SV | 14±6 | 42 | 0.6 | [104] |
| W | 4.2 | CPP-S/SV | 4.8±1 | 92±10 | 0.4 | [48] |

\#. For WL samples, $\rho_o = (RWt)/L$ includes surface scattering, as the films are thin—typically t ~ 20-50 nm.
\*\*. Corrected misprint of 75 for Ag in [97]

**IIIC. Table III. Spin-Diffusion Lengths, $l_{sf}^F$, in 'Nominally Pure' and Alloyed Ferromagnetic Metals**.

Values of $l_{sf}^F$ can be derived in two different ways from CPP-MR measurements with VF theory: (a) from variations of AΔR with $t_F$ using CPP-S/SVs (Section IIB2.b1 ), or (b) from values of MR vs $t_F$ in CPP-NW (Section IIB2.c). In principle, values of $l_{sf}^F$ can also be derived from measurements of the magnitude of ΔR in lateral transport, but, we feel that such comparisons of absolute magnitudes with theories are usually less certain, for the reasons discussed in Section II.C.4. To test for approximate proportionality of $l_{sf}^F$ to λ, Fig. 17 shows a plot of values of $l_{sf}^F$ vs $1/\rho_F$ from CPP-MR measurements at 4.2K. The $l_{sf}^{Co}$ shown in the inset is anomalously long. This may well be because $\rho_{Co}$ is dominated by scattering from stacking faults, which might flip electron spins only weakly. But, as explained in sections IIB2.b1 and IIB2.c1, there is a small possibility that the inferred values of $l_{sf}^{Co}$ are too long. The value of $l_{sf}^{Fe}$ is put in the inset because otherwise it fell in the middle of the sample listing on the figure. The straight line fit neglects the Co and Fe points.

| Metal | T(K) | Technique | $l_{sf}^F$ (nm) | $\rho_o$(nΩm) | $\rho_o l_{sf}^F$ (fΩm$^2$) | Ref. |
|---|---|---|---|---|---|---|
| Co | 77 | CPP-NW | 59±18 | 160±20 | 9 | [61] |
| Co | 300 | CPP-NW | 38±12 | 210±30 | 8 | [61] |
| Co | 4.2 | CPP-S/SV | ≥ 40 | 60 | ≥2.4 | [47] |
| Fe | 4.2 | CPP-S/SV | 8.5±1.5 | 40 | 0.34 | [106] |
| Ni | 4.2 | CPP-S/SV | 21±2 | 33±3 | 0.7 | [107] |
| Py = Ni$_{84}$Fe$_{16}$ | 4.2 | CPP-S/SV | 5.5 ±1 | 120 | 0.7 | [13] |
| Py | 77 | CPP-NW | 4.3±1 | | | [25] |
| Py | 293 | LNL/+ | 3 | 278 | 0.8 | [84] |
| Py | 79 | LNL/+ | 14.5 | 236 | 3.4 | [86] |
| Ni$_{66}$Fe$_{13}$Co$_{21}$ | 4.2 | CPP-S/SV | 5.5 ±1* | 90 | 0.5 | [108] |
| Co$_{91}$Fe$_9$ | 4.2 | CPP-S/SV | 12 ± 1 | 70 | 0.8 | [46] |
| Ni$_{93}$Cr$_3$ | 4.2 | CPP-S/SV | 3 ± 1 | 230 | 0.7 | [109] |

\* The value of $l_{sf}$ for Ni$_{66}$Fe$_{13}$Co$_{21}$ was not derived from a detailed fit to the data, but rather assumed from comparison of the data with those for Ni$_{84}$Fe$_{16}$.

### IIID. Table IV. Spin-Flipping Parameters, $\delta_{N1/N2}$, at N1/N2 Interfaces at 4.2K.

Table IV contains published values of spin-flipping parameters, $\delta_{N1/N2}$, for N1/N2 Interfaces at 4.2K, determined using the technique described in Section IIB2.b2. The probability $P$ of spin-flipping at each interface is $P = [1 - \exp(-\delta)]$. In hopes of elucidating the physics involved, the values are ordered from smallest to largest. Note that, for Cu, $\delta_{N1/N2}$ is smallest when paired with low atomic number metals V and Ag, and largest when paired with high atomic number metals Pt and W. Note also that there is no obvious correlation between $\delta_{N1/N2}$ and $2AR_{N1/N2}$. Caveats about how fundamental these values are, and brief remarks about the few published results for F/N interfaces, are given in Section IIB2.b2.b. We do not place the F/N results also in a table because we view most of them as highly uncertain.

| Metals (N1/N2) | T(K) | Technique | $\delta_{N1/N2}$ | $2AR_{N1/N2}$ (f$\Omega$m$^2$) | Ref. |
|---|---|---|---|---|---|
| Ag/Cu | 4.2K | CPP-S/SV | ~ 0 | 0.1 | [48] |
| V/Cu | 4.2K | CPP-S/SV | 0.07 ± 0.04 | 2.3 | [48] |
| Pd/Au | 4.2K | CPP-S/SV | 0.08 ± 0.08 | 0.45 | [30] |
| Au/Cu | 4.2K | CPP-S/SV | $0.13^{+0.08}_{-0.02}$ | 0.3 | [98] |
| Pt/Pd | 4.2K | CPP-S/SV | 0.13 ± 0.08 | 0.28 | [31] |
| Pd/Ag | 4.2K | CPP-S/SV | 0.15 ± 0.08 | 0.7 | [30] |
| Nb/Cu | 4.2K | CPP-S/SV | 0.19 ± 0.05 | 2.2 | [48] |
| Pd/Cu | 4.2K | CPP-S/SV | $0.24^{+0.1}_{-0.05}$ | 0.9 | [104] |
| Ru/Cu | 4.2K | CPP-S/SV | ~ 0.35 | 2.2 | [105] |
| Pt/Cu | 4.2K | CPP-S/SV | 0.9 ± 0.1 | 1.5 | [104] |
| W/Cu | 4.2K | CPP-S/SV | 0.96 ± 0.1 | 3.1 | [48] |

### IV. Summary and Conclusions.
### IVA. Summary of Results.

Table I. Measured values of $l_{sf}^N$ for Cu- or Ag-based alloys in which scattering is dominated by known concentrations of known impurities agree remarkably well with values calculated from the independently measured spin-orbit cross-sections or from analysis of spin-spin scattering. Since the dominant impurity in each alloy, and its concentration, are known, we expect $l_{sf}^N$ to be an intrinsic property of the alloy. The agreement between experimental and calculated values supports this expectation. This agreement also leads us to conclude that the Valet-Fert (VF) theory provides a good basis for evaluating CPP-MR data, and that the two independent techniques used in these measurements both seem to be valid at the 10%-20% level. Table I also shows that different impurities can have very different spin-orbit or spin-spin cross-sections, leading to very different values of $l_{sf}^N$ for a given impurity concentration.

Table II. Measured values of $l_{sf}^N$ are listed for a variety of nominally 'pure' metals at temperatures ranging from 4.2K to 293K. In Figs. 14 (Cu), 15 (Ag), and 16 (Al), these values are plotted against inverse resistivity ($1/\rho$) to see if they are proportional to the transport mean-free-path, $\lambda_t$. In a few cases, there is apparent rough scaling, but such scaling is not general. At cryogenic temperatures, the resistivity of such metals is dominated by scattering from an unknown concentration of unknown impurities. Thus $l_{sf}^N$ is not intrinsic, but is essentially unique to each sample, and need not grow linearly with total impurity content (roughly measured by the residual resistivity, $\rho_o$). For a sufficiently high purity metal, where phonon scattering is dominant, one might expect $l_{sf}^N$ at 293K to be intrinsic. There is no evidence of 'limiting high purity' values of $l_{sf}^N$ at 293K for the samples in Table II or in Figs. 14-16, where we list in each caption the value of $1/\rho_N$ where this limiting value of $l_{sf}^N$ would be expected.

Table III. Measured values of $l_{sf}^F$ are listed for several nominally 'pure' and alloyed F-metals, mostly at 4.2K. Except for Co, the values are all ≤ 20 nm. Fig. 18 shows that some of the 4.2K values correlate with ($1/\rho_F$).

Table IV. Measured values of $\delta_{N1/N2}$ are listed for several metal pairs. The values show some correlation with difference in atomic number, as expected from simple spin-orbit arguments, but no particular correlation with interface specific resistance, $AR_{N1/N2}$.

In section IIB2.b2.b, we describe several inferences of non-zero values of $\delta_{F/N}$, most of which we view as highly uncertain.

### IVB. Advantages and Disadvantages of Different Measuring Techniques.

**Current-Perpendicular to Plane MR with Superconducting Cross-Strips (CPP-MR/S).**

Advantages: The geometry is well controlled, crucial parameters can be measured independently, and certain techniques seem to have been well validated for determining $l_{sf}^N$, $l_{sf}^F$, and $\delta_{N1/N2}$.

Disadvantages: So far, measurements have been made only at T = 4.2K, and the technique has been used only for $l_{sf}^N$, $l_{sf}^F \leq 100$nm. However, bulk and interface asymmetry parameters and interface specific resistances might be only moderately sensitive to temperature [57].

**Current-Perpendicular-to-Plane MR with Nanowires (CPP-MR/NW).**

Advantages: In principle, $l_{sf}^N$ and $l_{sf}^F$ can be measured from below T = 4.2K to above T = 293K. $l_{sf}^F$ can be determined from a straight line plot when $t_F \gg l_{sf}^F$ and $t_N \ll l_{sf}^N$ (Eq. 18).

Disadvantages: Pure N- and F-layers are difficult to obtain; contamination of the F-layers can be particularly severe. Determining $l_{sf}^N$ requires a numerical fit.

**Lateral-Non-Local (LNL) Measurements.**

Advantages: Long $l_{sf}^N$ can be measured from below T = 4.2K to above T = 293K. So far, this technique has been used to measure $l_{sf}^N$ only in Ag, Al, Cu, and (with less certainty) Au. In principle, one can infer $l_{sf}^F$, but less directly. The two published LNL estimates of $l_{sf}^{Py}$, obtained in different ways, differ by almost a factor of five.

Disadvantages: Indirect determination of $l_{sf}^F$ requires knowing several experimental parameters (see Eq. 16a). To get uniform spin-current, sample width must be much less than sample length ($W \ll L$). To use a simple single-exponential equation, low resistance F/N contacts require $L > 2 l_{sf}^N$, which can give weak signals. Combined, these constraints mean that to measure short $l_{sf}^N$ will require narrow N-films.

**Weak Localization (WL)**

Advantages: $l_{so} = l_{sf}^N$ can be measured reliably at T $\leq$ 40K.

Disadvantages: To separate effects of $l_{so}$ and $\lambda_\phi$, requires a measuring T such that $\lambda_\phi$ is comparable to $l_{so}$. This requirement is not necessarily too-stringent, since $\lambda_\phi$ varies rapidly with temperature.

### IVC. Some Needs for Additional Work.

(1) Badly needed is a direct technique for measuring $\delta_{F/N}$.
(2) Badly needed are calculations of $\delta_{N1/N2}$ and $\delta_{F/N}$, especially to establish whether there is a large difference for 'perfect' versus 'alloyed' interfaces.
(3) A different way to measure $\delta_{N1/N2}$ would be useful, to independently check the values in Table IV.
(4) The ability to produce narrower structures (W $\leq$ 30 nm) should allow reliable LNL measurements to be extended to metals with much smaller values of $l_{sf}^N$ than can be studied with the W $\geq$ 200 nm N-films of most published measurements.

**Acknowledgments:** We thank the following for helpful comments, clarifications, corrections, and suggestions for improvement: N.O. Birge, A. Fert, G.E.W. Bauer, B. Doudin, P.M. Levy, M. Johnson, M.D. Stiles, R.A. Webb, W.H. Butler, S. Takahashi, A. Manchon, and B.J. van Wees. We, of course, take full responsibility for any remaining errors and unclarities. We also thank the US NSF for support under grant DMR 05-01013.

## Appendix A. Definitions of Parameters and Spin-Accumulation Equations in terms of $l_{sf}^N$ and $l_{sf}^F$.

### A1. Parameters and relationships.

Within the F-metal we define parameters [7, 8]

$$\rho_F^* = (\rho_F^\downarrow + \rho_F^\uparrow)/4 \tag{A1}$$

and

$$\beta_F = (\rho_F^\downarrow - \rho_F^\uparrow)/(\rho_F^\downarrow + \rho_F^\uparrow) \tag{A2}$$

Additivity of conductivities for simple transport gives

$$\sigma_F = \sigma_F^\downarrow + \sigma_F^\uparrow. \tag{A3}$$

From $\sigma = (1/\rho)$, we get $(1/\rho_F) = (1/\rho_F^\downarrow) + (1/\rho_F^\uparrow)$, which can be rearranged to give

$$\rho_F = \rho_F^* (1 - \beta_F^2) \tag{A4}$$

Eq. (A4) relates the parameter $\rho_F^*$ to $\rho_F$, the separately measured resistivity of a thin film of F. Note that replacing F by N with $\sigma_N^\downarrow = \sigma_N^\uparrow = \sigma_N/2$, gives $\beta = 0$ and $\rho_N = \rho_N^*$, as required.

Now we turn to multilayers. To simplify, we consider a one-dimensional multilayer, involving just a single F-metal and a single N-metal, with the direction z along the sample CPP axis. We let both F and N have the same free-electron Fermi surface, but different conductivities, $\sigma_N$ for N and $\sigma_F^{\downarrow,\uparrow}$ for F as above, and different elastic scattering times, $\tau^N$ for N, and $\tau_{\uparrow,\downarrow}^F$ for F, leading to mean-free-paths,

$$\lambda^N = v_F(1/\tau^N + 1/\tau_{sf}^N)^{-1} \tag{A5}$$

and

$$\lambda_{\uparrow,\downarrow}^F = v_F (1/\tau_{\uparrow,\downarrow}^F + 1/\tau_{sf}^F)^{-1}, \tag{A6}$$

with different spin-relaxation times, $\tau_{sf}^N$ for N and $\tau_{sf}^F$ for F. In F,

$$l_{\uparrow,\downarrow}^F = \sqrt{(1/3)v_F \lambda_{\uparrow,\downarrow}^F \tau_{sf}^F} \tag{A7}$$

and $l_{sf}^F$ is given by [8, 25]

$$(1/l_{sf}^F)^2 = (1/l_\uparrow^F)^2 + (1/l_\downarrow^F)^2 \tag{A8}$$

Following [25], we insert Eqs. A7 into A8 and solve for $l_{sf}^F$, finding

$$l_{sf}^F = \sqrt{(\lambda^{F*} \lambda_{sf}^F)/6}, \tag{A9}$$

where

$$(1/\lambda^{F*}) = (1/2)[(1/\lambda_\uparrow^F) + (1/\lambda_\downarrow^F)]. \tag{A10}$$

In the free-electron model, we assume that each spin-channel in F contains half of the electrons. Converting Eq. A3 from $\sigma$ to $\lambda$ thus gives

$$\lambda_t^F = (1/2)(\lambda_\downarrow^F + \lambda_\uparrow^F). \tag{A11}$$

From the definition of $\beta_F$ in terms of $\rho^\downarrow$ and $\rho^\uparrow$, and the inverse relation between $\rho$ and $\lambda$, we can take $\lambda^\uparrow = (1+\beta_F) \lambda_t^F$ and $\lambda^\downarrow = (1-\beta_F) \lambda_t^F$, and Eq. A10 gives

$$\lambda^{F*} = \lambda_t^F (1-\beta_F^2). \tag{A12}$$

Eq. (A9) thus becomes Eq. (2).

Combining Eqs. 5, A4 and A12 gives

$$\lambda^{F*} \rho_F^* = \lambda_t^F \rho_F = \rho_b l_b, \tag{A13}$$

where $\rho_b l_b$ is defined in Appendix B.

Finally, to obtain the appropriate equations for N, we simply let ↑ parameters = ↓ parameters. Then $\beta_N = 0$, $\lambda^{N*} = \lambda_t^N = \lambda_\uparrow^N = \lambda_\downarrow^N$, $\rho_N^* = \rho_N$, and Eq. A9 becomes Eq. (3).

$$l_{sf}^N = \sqrt{(\lambda_t^N \lambda_{sf}^N)/6}. \tag{3}$$

**A2. Equations for spin-accumulation, $\Delta\mu$, in terms of $l_{sf}^F$ and $l_{sf}^N$.**

In each layer, define current densities, $j_{\uparrow,\downarrow}$, and chemical potentials, $\mu_{\uparrow,\downarrow}$. In the limit $\lambda << l_{sf}$, the equations governing electron transport in the F-layers are [8, 110-112]:

Ohm's law: $\quad\quad\quad\quad \partial\mu_{\uparrow,\downarrow}^F/\partial z = (e/\sigma_{\uparrow,\downarrow}^F) j_{\uparrow,\downarrow}^F \tag{A14}$

and

Diffusion Equation: $\quad\quad\quad \partial^2(\mu_\uparrow^F - \mu_\downarrow^F)/\partial z^2 = (\mu_\uparrow^F - \mu_\downarrow^F)/(l_{sf}^F)^2 \tag{A15}$

Eq. A14 is just Ohm's law for each spin-direction. If we define the 'spin-accumulation' $\Delta\mu = (\mu_\uparrow^F - \mu_\downarrow^F)$, Eq. A15 is a diffusion equation for $\Delta\mu$, with scaling length $l_{sf}^F$. The solution to Eq. A15 in one-dimension is:

$$\Delta\mu = A\exp(-z/l_{sf}^F) + B\exp(z/l_{sf}^F) \tag{A16}$$

In a free-electron model, $|\Delta\mu| = 2\mu_o |\Delta M|/(3n\mu_B)$ is related to the out of equilibrium magnetization, $\Delta M$, where n is the electron density, $\mu_B$ is the Bohr Magneton, and $\mu_o$ is the magnetic permeability of empty space. Eq. A16 then says that the out of equilibrium magnetization can grow or decay exponentially with length $l_{sf}^F$. This direct proportionality between $\Delta\mu$ and $\Delta M$ is the source of the term 'spin-accumulation'—i.e., $\Delta\mu \neq 0$ means that non-equilibrium spins (magnetic moments) build up or decay in the sample. The details of how they do so in a given multilayer structure are determined by the VF equations [8]. In general, $\Delta\mu(z)$ in an F-layer includes both terms in Eq. A16.

For N-layers, the governing equations are:

Ohm's law: $\quad\quad\quad\quad \partial\mu_{\uparrow,\downarrow}^N/\partial z = (e/\sigma^N) j_{\uparrow,\downarrow}^N \tag{A17}$

and

Diffusion Equation $\quad\quad\quad \partial^2(\mu_\uparrow^N - \mu_\downarrow^N)/\partial z^2 = (\mu_\uparrow^N - \mu_\downarrow^N)/(l_{sf}^N)^2. \tag{A18}$

The 1D solution to Eq. A18 is just Eq. A16 with $l_{sf}^F$ replaced by $l_{sf}^N$. In general, as in F-layers, both terms in Eq. A16 must be included for each N-layer. But careful experimental design, as in most of the experiments described in Section II.B and some in Section II.C, can leave only the decaying exponential.

Cautionary Note: One must examine details when comparing our analysis with those in other papers, since chosen relationships can differ from ours by factors of two—e.g. some choose $2\Delta\mu = (\mu_\uparrow^F - \mu_\downarrow^F)$; $(1/\lambda^{F*}) = [(1/\lambda_\uparrow^F) + (1/\lambda_\downarrow^F)]$; etc. Hopefully, properly interpreted, the final results turn out to be the same.

**Appendix B. Defining $\lambda_t$ for a metal with a measured resistivity, $\rho$.**

For a cubic non-magnetic metal, the electrical conductivity $\sigma_E$ can be written as an integral of the mean-free-path $\lambda$ over the area of the Fermi surface $S_F$ [113]:

$$\sigma_E = (e^2/12\pi^3 \hbar) \int \lambda dS'_F . \tag{B1}$$

Here e is the electronic charge, $\hbar$ is Planck's constant divided by $2\pi$, and $\lambda$ is to be integrated over $S_F$.

If $\lambda = \lambda_t$ is constant over $S_F$, it can be removed from the integral and, using $\sigma_E = 1/\rho$, Eq. B1 can be rewritten as:

$$\lambda_t = [(12\pi^3 \hbar/e^2 S_F)]/\rho \equiv \rho_b l_b /\rho, \tag{B2}$$

which defines the constant $\rho_b l_b$ in Eq. 5. For a free-electron gas, $\rho_b l_b$ can be written in several different ways. Eq. 2.91 in [39] gives

$$\lambda_t = (r_s/a_o)^2 (9.2 \text{ nm})/\rho, \tag{B3}$$

where $a_o = \hbar^2/me^2$ is the Bohr radius and $r_s$ is the radius of a sphere whose volume is the volume per conduction electron. Values of $r_c/a_o$ for several metals are listed in [39], from which values of $\rho_b l_b$ range from 0.4 f$\Omega$m$^2$ for Al to 2.9 f$\Omega$m$^2$ for Cs.

Using the quantum of resistance, $R_q = h/e^2 \approx 26$ k$\Omega$, and the Fermi wavevector $k_F$ with $4\pi k_F^2 = S_F$, Eq. (B2) can be rewritten as

$$\lambda_t = [(R_q)(3\pi)/2k_F^2]/\rho. \tag{B4}$$

Values for $\rho_b l_b$ can be determined by: (a) calculation for either free electrons or using a real Fermi surface; (b) anomalous skin effect (ASE) studies; and (c) size effect studies on thin wires or films. A collection of values from all three is given in [40]. For Cu, the free-electron and ASE values are similar and agree with most of the size-effect studies listed. For Ag, the ASE values are 40% higher than the free-electron calculation, and the size-effect studies agree slightly better with the free-electron calculation.

Given the agreement between the free-electron, ASE, and size-effect results just described, plus the surprisingly close agreements between the measured spin-diffusion lengths in Cu- and Ag-based alloys given in Table I and those calculated using free-electron values of $\rho_b l_b$ for Cu and Ag, we infer that free-electron values of $\lambda_t$ for Cu and Ag should be reliable to at least 50% and perhaps better. This conclusion disagrees with a claim [114] that the values of $\lambda_t$ for Cu might be in error by a factor of five.

**Appendix C. Consideration of Mean-Free-Path Effects in CPP Transport.**

We noted above that the 2CSR model used to analyze the CPP-MR, and the VF and related models used to analyze both CPP-MR and LNL data, were derived assuming free-electron Fermi surfaces. Because the 2CSR model is so simple, it provides a convenient baseline for comparing with calculations that take account of real Fermi surfaces and, indeed, as we'll see in item (4) below, the resulting values of the 2CSR parameters can agree with calculations that include band structure effects. Nonetheless, for several years, theorists have been showing that including real-Fermi surfaces could lead to deviations from the 2CSR model, even with no spin-flip scattering. Calculations have examined effects of ballistic versus diffuse scattering within the F- and N-layers [35], as well as of perfect (ballistic scattering) verse intermixed (diffusive scattering) interfaces [115]. Deviations from the 2CSR model can arise from: (a) quantum coherent effects, such as quantum well states [32, 36], and (b) electronic distribution functions in the Boltzmann equation that vary exponentially on the scale of the mean-free-path in the vicinity of interfaces [33, 34, 115, 116]. Probably the simplest way to summarize these analyses is that they predict that interfacial specific resistances (AR) can depend upon the separation of the interfaces when that separation is comparable to or less than a mean-free-path ($\lambda$). They are, thus, called mean-free-path (mfp) effects.

In general, the predicted deviations from the 2CSR model are largest for ballistic transport in the N- or F-metals and for perfect interfaces, in part because ballistic transport and perfect interfaces enhance quantum coherence. In practice, real interfaces are not perfect, and contributions to the 2CSR model from these imperfect interfaces can dominate over the bulk, especially if quasi-ballistic bulk transport makes the bulk contributions small. In such a case, it is unclear how large any deviations from the 2CSR model might be, and in our view the best way to clarify the situation is via experiments.

Indeed, a number of experiments have been made to explicitly test the 2CSR model and the VF extension to include spin-flipping. In this Appendix we outline the results obtained, considering tests of both 2CSR and VF. This distinction is important, because two coupled groups [114, 117-119] have argued that a number of observed deviations from the 2CSR should be attributed not to effects of finite $l_{sf}$ via the VF model, but rather to mfp-effects (although we shall see that the two groups apply different mfp-models). We consider several experiments sequentially.

(1) In the first test of the 2CSR model [7], Eq. 7b was applied to data on Ag and Ag(4%Sn) alloys, with Sn chosen as giving a large increase in residual resistivity per atomic percent impurity, but having only a small spin-orbit interaction (i.e., weak spin-flipping). As shown in Fig. 3 and Table I, the data for Ag and Ag(4%Sn) fell closely on the same straight line through the origin, although their residual resistivities (and thus their mean-free-paths) differ by about a factor of twenty. This result was confirmed using Cu and Cu(4%Ge) (Fig. 4 and Table I), since Ge in Cu also gives strong elastic scattering, but weak spin-flipping [26, 40]. As additional evidence that the resistance of a Co/Ag interface does not depend upon the thickness $t_{Ag}$ of the separating Ag layer, Fig. 18 shows that data for fixed $t_{Co}$ = 6 nm plotted as in Eq. 7b, fall on the same straight line passing through the origin when the Ag thickness is held fixed at $t_{Ag}$ = 6 nm or let vary from $t_{Ag}$ = 12 nm to 36 nm.

(2) The first quantitative test of the VF model involved a series of Ag- and Cu-based alloys with impurities (first Pt [12] and later Ni [94]) having spin-orbit cross-sections known from CESR measurements [26] to be large enough to give noticeable deviations from Eq. 2b. Figs. 3 and 4 show several examples of such data. The resulting values of $l_{sf}^N$ given in Table I agree well with predictions from the CESR results.

To summarize, Figs. 3 and 4 show two different behaviors: (a) when spin-flipping is weak, data for high resistivity AgSn and CuGe agree with data for Ag and Cu and the 2CSR model, but (b) when spin-flipping is strong, data for lower resistivity AgPt, CuPt, and CuNi disagree with the 2CSR model but are quantitatively explained by the VF model. No-one has yet explained both (a) and (b) based solely upon mfp effects.

(3) Mfp-effects have also been proposed [32, 119] as an alternative explanation to finite $l_{sf}^N$ for the different variations of the slopes of a plot of log (A$\Delta$R) vs $t_N$ in Fig. 9 for different nominally pure metals. However, the slopes don't correlate with the residual resistivities (nominal mean-free-paths) of the metals, but do correlate with the weights of the metals (light V, medium Nb, and heavy W), as expected if $l_{sf}^N$ is dominated by spin-orbit scattering due to interfacial alloying with Cu.

(4) Next, if the interface resistances, AR, change with layer thickness for layers that are thinner than their mean-free-paths, one would expect interface resistances in [N(3)/Cu(3)]$_N$ multilayers to be affected, because the mfp at 4.2K for the sputtered Cu is ~ 120 nm using the parameters given in the text above. However, in all four cases studied so far of [M1/M2] where M1 and M2 have the same crystal structure and the same lattice parameter to within 1%, values of AR calculated with no adjustable parameters, and including real band structures, agree with published experimental values to within their mutual uncertainties [30, 120]. The calculations assumed diffusive scattering within the layers and found only modest differences between values of AR for interfaces that were perfect or else two-monolayers of a 50%-50% alloy. Again, there is no evidence of significant deviations from 2CSR or VF analyses that would require the presence of mfp effects.

(5) The strongest case for deviations from 2CSR model behavior comes from comparisons of data on [F1/N/F2/N] multilayers in which the two different layers F1 and F2 are interleaved—[F1/N/F2/N]$_N$, or separated [F1/N]$_N$.[F2/N]$_N$. The first experimental evidence of a potential problem with the 2CSR and VF models was a report [117] that, contrary to expectations from the 2CSR model, A$\Delta$R at 4.2K differed for samples of Co/Cu that were 'interleaved'--[Co(8)/Cu(10)/Co(1)/Cu(10)]$_N$, or 'separated'—[Co(8)/Cu(10)]$_N$[Co(1)/Cu(10)]$_N$, where Co(8) and Co(1) represent Co layers of thicknesses 8 nm and 1 nm that have different reversing fields, and the Cu layers are thick enough to minimize magnetic coupling between adjacent Co layers. Similar differences had previously been reported in [Co, Py] [121] and [Co, Fe] [122] multilayers, but there can plausibly be attributed to spin-flipping within the Py or Fe layers due to short spin-diffusion lengths (see Table III). For Co/Cu, in contrast, the best estimates of $l_{sf}^F$ (Table III) are much larger than the layer thickness $t_{Co}$ = 8 nm or 1 nm. The authors of [117] suggested two independent explanations for the differences: (a) mfp effects due to real Fermi surface effects as described by Tsymbal and Pettifor [32], or (b) a phenomenological approach in which AR depends upon only the relative orientations of adjacent Co layers. For simplicity, we characterize the disagreements in both cases as involving mfp effects versus spin-flipping.

(a) The data of [117] were confirmed in [16, 52], so there is no issue about their correctness. The issue is solely their interpretation. To test the Tsymbal argument, the measurements were repeated replacing the Cu by Cu(2%Ge), which has a mfp about 15 times shorter than Cu, but is only a weak spin-flipper [26]. The relative differences in A$\Delta$R for interleaved and separated multilayers were unchanged—the differences are insensitive to mean-free-path in this regime. Subsequently, replacing Co by CoFe and Py revealed similar relative differences [123]. Also, inserting into just the central Cu layer of a separated [Co(8)/Cu]$_N$[Co(1)/Cu]$_N$ multilayer a 2nm thick layer of the strong spin-flipper FeMn increased the difference from the interleaved one, even though the total AR of the central layer including the FeMn did not exceed that of a full layer of CuGe [123]. These results were taken [123] as evidence that spin-flipping, rather than mfp effects, was the source of the differences. If one accepts a long $l_{sf}^{Co}$, then such spin-flipping in the simple Co/Cu multilayers must occur at the Co/Cu interface. [52] argued that the observed differences could be explained by a $\delta_{Co/Cu}$ = 0.25. Later, that same value of $\delta_{Co/Cu}$ was shown [16] to improve the predictions of A$\Delta$R for Co/Cu EBSVs based upon the Co/Cu parameters previously determined from multilayer studies, and to help account for slower than expected CPP-MR growth when 'internal interfaces with Cu' were inserted within Co layers [53]. So far, however, there is no independent confirmation of such a $\delta_{Co/Cu}$.

(b) The phenomenological approach is based upon the assumption that an electron is only weakly scattered as it travels through the N-layer, so that "one must consider the electron as being scattered by the combined potential of a pair of neighboring F-layers" [114]. Without providing any more fundamental justification than this simple statement, the authors developed a simple model, with adjustable parameters, in which AR depends only upon the angle between adjacent Co layers, They showed that they could fit various sets of their data. Although the authors call this a mean-free-path model, it contains no characteristic lengths—i.e., no mean-free-paths appear in it. Contrary to the evidence above that the 2CSR model works well for samples with $t_N \leq \lambda$, these authors claim that a CPP-MR review [5] specifies a requirement for 2CSR applicability to be t $\gg \lambda$. But examination of that review shows that it is more careful. Consistent with a statement on page 302 of that review, the theorist co-author has allowed us to say the following: "The important $\lambda$ is definitely not the bulk mean free path, $\lambda_b$, which for thin multilayers is not very relevant. Rather, the two channel resistor model is often relevant because of the diffuse scattering at the interfaces. In that case you still can work with a mean-free-number of transmitted interfaces **N**. In a very simplified picture, where d is the repeat period, you might say that $1/\lambda = 1/(Nd) + 1/(\lambda_b)$." Here **N** is the mean-number of interfaces through which an electron passes before being scattered, typically **N** ~ 2. Intriguingly, the latest data from this group [114] confirm the counter-argument that spin-flipping produces differences between interleaved and separated samples similar to those of interest.

These authors also claimed [118] that $\lambda$ was calculated inappropriately in [16], because $\rho_b l_b$ for Cu might be 5 times larger than estimated. Counter-arguments are given in Section II.A and Appendix B.

(6) Lastly, a more recent, independent test of the 2CSR model and mfp effects used EBSVs of Co/Cu, Co/Ag, and Co/Au [124]. The square root function of Eq. 7b was examined, holding $t_{Co}$ fixed, and varying only $t_N$. Eq. 7b of the 2CSR model predicts that the square root should stay constant, independent of $t_N$. In contrast, both mfp effects and the VF equations predict that the square root should decrease with increasing $t_N$, but for very different reasons. MFP effects cause a decrease because the interface resistance changes as $t_N$ increases. This decrease should 'saturate' to a constant value for $t_N$ beyond a certain value. In contrast, VF predicts a decrease if the ratio $t_N/l_{sf}^N$ is large enough to cause significant spin-flipping. Here, the deviation should increase indefinitely with increasing $t_N$. Unfortunately, the range of thicknesses studied was not large enough to look for this difference. The observed deviations from constancy were consistent with simple linear variations with $t_N$. The results of the experiments were interpreted differently by two groups. One [124] argued that the observed decreases of the Ag- and Au-based data were consistent with the values of $l_{sf}^N$ given in Table II above, so there was no need for mfp effects. While the decreases of the Cu-based data were larger than expected from the values of $l_{sf}^N$ in Table II, and could be understood based upon mfp effects, the best fit difference between the two models was only about one standard deviation, too small to claim an unambiguous mfp effect. They conceded, however, that the uncertainties in all three cases were sufficient that a modest mfp effect could not be ruled out. The other [119] showed that mfp effects could fit the Co/Cu data and claimed those data to be evidence of such effects. About these data and mfp effects, we make the following points. (1) For Co/Cu, the values of both AR(AP) and AR(P) were correctly obtained from the 2CSR model with previously determined parameters with no adjustments. So, even in the best case for mfp effects, such effects were not needed to explain AR(AP) and AR(P). Their possible effect could be seen only in the discrepancy from the 2CSR model in the square root data, which depends upon small differences between AR(AP) and AR(P). (2) In contrast, the mfp fits to the data are not parameter free, but involve both a parameter for the square root and a second parameter to get AR(AP) and AR(P) approximately correct. (3) Since no mfp calculations have yet been made for Co/Ag and Co/Au, any need for mfp-effects there is unclear.

To conclude, the strongest evidence for some mfp effects is the difference in behaviors of AR for interleaved [Co(8)/Cu/Co(1)/Cu]$_N$ and separated [Co(8)/Cu]$_N$[Co(1)/Cu]$_N$ multilayers. If a long $l_{sf}^{Co} \sim 60$ nm at 4.2K is accepted, then this difference seems to require either some mfp effects or else spin-flipping at Co/Cu interfaces. The rest of the information in this Appendix gives us the impression that, with this possible exception, any mfp-effects are rarely if ever beyond experimental uncertainty ~ 10-20%. Since there is disagreement between different groups over this issue, the reader will have to make his or her own judgment.

Figure Captions. (Format with J. Non-C. S.)

Fig. 1. CPP $AR_T(H)$ vs H for: (a) a simple $[Co(6)/Ag(6)]_6$ multilayer with all Co-layers having equal thickness; (b) a $[Co(8)/Cu(10)/Co(1)/Cu(10)]_4$ multilayer with F-layers of alternating thicknesses, and (c) an EBSV of the form [FeMn(8)/Py(24)/Cu(10)/Py(24)]. Cases (b) and (c) give stable values of AR(AP) that reproduce during multiple field sweeps when the sample is taken to high-field saturation. In case (a), in contrast, the maximum AR obtained after saturation is AR(Peak), but the best estimate of AR(AP) is AR(0), an initial state that does not reproduce under field sweeps. Demagnetization of a simple multilayer usually gives values of AR(H) between AR(0) and AR(Peak) [38] .

Fig. 2. CPP-MR Geometries. (a) Superconducting Cross-strips with short-wide sample ($L << W$); (b) nanowires ($L >> W$); (c) nanopillars, $L \sim W$.

Fig. 3. Fig. 3. $\sqrt{A\Delta R(AR(AP))}$ vs $N$ for Ag and Ag-based alloys. The numbers to the right of the curves indicate $l_{sf}^N$ = 11 nm for Ag(6%Mn), 10 nm for Ag(6%Pt), and 7 nm for Ag(9%Mn). From [12]

Fig. 4. $\sqrt{A\Delta R(AR(AP))}$ vs $N$ for Cu and Cu-based alloys. The numbers to the right of the curves indicate $l_{sf}^N$ = 8 nm for Cu(6%Pt) and 2.8 nm for Cu(7%Mn). After [12] and [45].

Fig. 5. $A\Delta R$ vs $t_{Py}$ for Py-based EBSVs. The solid curve is a fit to VF theory with $l_{sf}^{Py}$ = 5.5 nm. The dashed curve represents the expected variation for $l_{sf}^{Py}$ = ∞. Note that, for small $t_{Py}$, the solid curve lies above the dashed one, primarily because of the differences in the denominators of Eqs. 8 and 10. From [13]

Fig. 6. $A\Delta R$ vs $t_F$ for F = Co- and Co(9%Fe)-based EBSVs. The solid curve is a fit to VF theory with $l_{sf}^{CoFe}$ = 12 nm. The dashed line represents the expected variation for $l_{sf}^{CoFe}$ = ∞. Note that, for small $t_{CoFe}$, the solid curve lies above the dashed one, primarily because of the differences in the denominators of Eqs. 8 and 10. From [46].

Fig. 7. (a) Schematic of an EBSV wit a single X = N-layer insert. (b) Calculation of log ($A\Delta R$) vs $t_X$ for such an EBSV. The dashed line is Eq. 11 with a constant denominator. From [48].

Fig. 8. (a) Schematic of an EBSV with an X = $N_1(3)/N_2(1)]_N$ multilayer insert. (b) Calculated log ($A\Delta R$) vs $N$ for such an insert. The dashed line is Eq. 12 with a constant denominator. To simplify, in both cases we've assumed $l_{sf}^{N1} = l_{sf}^{N2}$ = ∞. From [48]

Fig. 9. log ($A\Delta R$) vs t for X = Ag, CuPt, V, Nb, W, and FeMn. With the exception of FeMn, where the curve is just a guide to the eye, the solid and dashed curves are fits to the VF theory with the parameters in Tables II and III. From [48].

Fig. 10. log ($A\Delta R$) vs $N$ for X = $[Cu/Ag]_N$, $[Cu/V]_N$, $[Cu/Nb]_N$, and $[Cu/W]_N$. The solid, broken, and dotted curves are fits using VF theory and the parameters in Tables II and III. The dashed curve indicates the expected behavior for $l_{sf}^{Cu/V}$ = ∞. From [48]

Fig. 11. R(P)/$\Delta R$ vs $t_{Co}$ at 77K and 295K. From [43].

Fig. 12. Lateral (L) Geometries for Standard and Non-local (NL) Measurements. (a) Lateral spin-valve film with Standard (#1) and non-local (NL) (#2) current and voltage connections. (b) LNL-Cross (LNL/C) geometry with F1 and F2 layers of different widths. (c) LNL/+ geometry with additional N2 and/or F3 cross-strips. (d) LNL/TTD three terminal device.

Fig. 13. $\Delta R$ vs $L$ for: (a) an LNL/T Co/I/Al/I/Co sample--● = 4.2K, o = 293K, data from Jedema et al., [73]; and (b) an LNL Py/Cu/Py samples-- ■ = 4.2K, data from Garzon [83], □ = 293K, data from Kimura et al. [84] After [82].

Fig. 14. $l_{sf}^{Cu}$ vs $1/\rho_{Cu}$ for Cu samples in Table II. References: Pierre [91]; Gougam [97]; Jedema [70, 77]; Takahashi [82]; Garzon [58]; Bergmann, [96], Ji [80]; Kimura [84]; Doudin [64]; Piraux [43]; Albert [65]. The line is a least-squares fit to the data for T ≤ 4.2K (filled symbols) constrained to go to (0,0) and neglecting the symbols (+ and x). Note: For pure Cu at 293K, $1/\rho_{Cu} = 0.060$ (nΩm)$^{-1}$ [40].

Fig. 15. $l_{sf}^{Ag}$ vs $1/\rho_{Ag}$ for Ag samples in Table II. Symbols: Pierre [91]; Gougan [97]; Bergmann [96]; Godfrey [86]. The solid line is a least-squares fit to the data for T < 2K (filled symbols) constrained to go to (0,0). The dashed line is a similar fit to the data for T = 79K and 298K. We omit from Fig. 15 the data point in Table II by Park et al. [48] which set only an extreme lower bound on $l_{sf}^{Ag}$. Note: For pure Ag at 293K, $1/\rho_{Ag} = 0.063$ (nΩm)$^{-1}$ [40].

Fig. 16. $l_{sf}^{Al}$ vs $1/\rho_{Al}$ for Al samples in Table II. Symbols: Jedema (02) [73]; Jedema (03) [77]; Johnson [3]; Santhanam [24, 99]; Wind [100]; Poli [88], Valenzuela [89]; Otani [90]. The straight line is a least-squares fit to the 4K data, constrained to go to (0,0). We use a log-log insert plot to place the higher purity samples of Johnson and Otani; the line in the insert is the same as in the main figure. Note: For pure Al at 293K, $1/\rho_{Al} = 0.038$ (nΩm)$^{-1}$ [40].

Fig. 17. $l_{sf}^{F}$ vs $1/\rho_{F}$ for CPP-MR samples in Table III. The main figure contains values for Ni and alloys, plus a best fit straight line to just those values and constrained to go to (0,0). The Insert contains this same line plus values for Co and Fe.

Fig. 18. $\sqrt{A\Delta R(AR(0))}$ vs $\mathcal{N}$ for Co/Ag with fixed $t_{Co}$ = 6nm comparing data for fixed $t_{Ag}$ = 6 nm and for fixed $t_T$ = 720 nm. The dashed line is a fit to the open circles passing through (0,0). After [45]

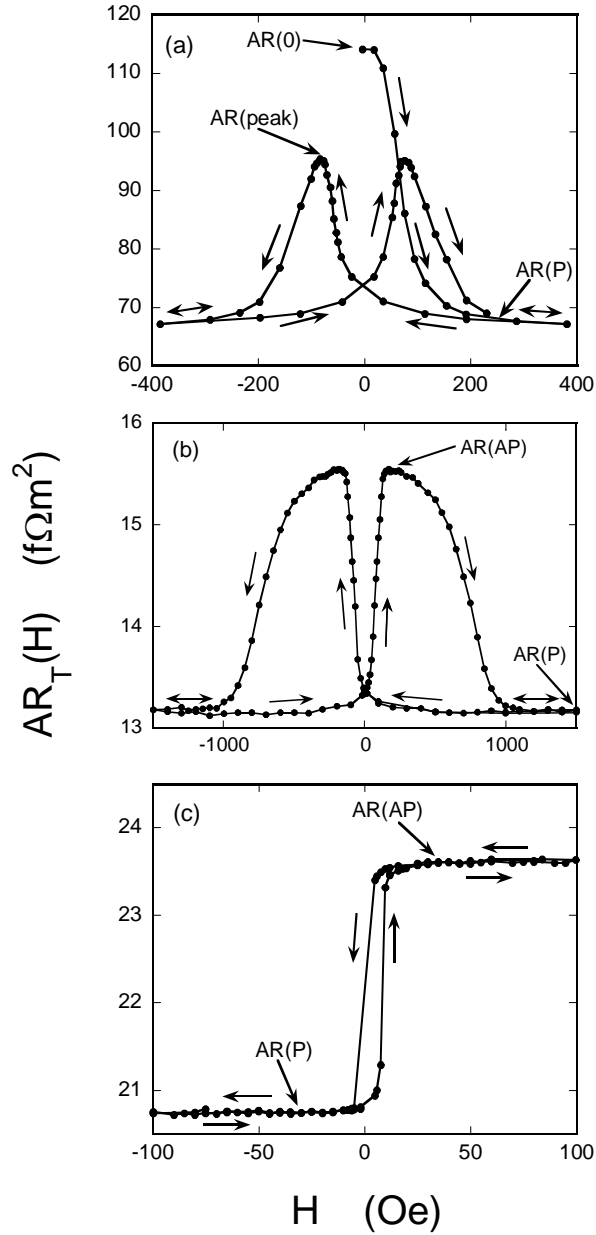

Fig. 1. CPP $AR_T(H)$ vs H for: (a) a simple $[Co(6)/Ag(6)]_6$ multilayer with all Co-layers having equal thickness; (b) a $[Co(8)/Cu(10)/Co(1)/Cu(10)]_4$ multilayer with F-layers of alternating thicknesses, and (c) an EBSV of the form [FeMn(8)/Py(24)/Cu(10)/Py(24)]. Cases (b) and (c) give stable values of AR(AP) that reproduce during multiple field sweeps when the sample is taken to high-field saturation. In case (a), in contrast, the maximum AR obtained after saturation is AR(Peak), but the best estimate of AR(AP) is AR(0), an initial state that does not reproduce under field sweeps. Demagnetization of a simple multilayer usually gives values of AR(H) between AR(0) and AR(Peak) [38].

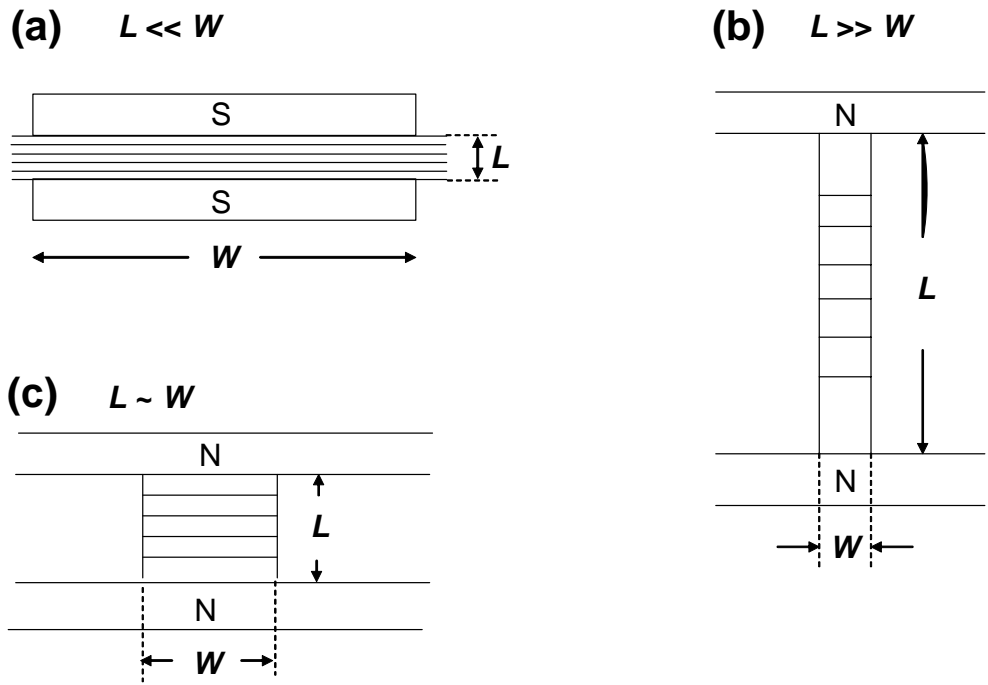

Fig. 2. CPP-MR Geometries. (a) Superconducting Cross-strips with short-wide sample ($L \ll W$); (b) nanowires ($L \gg W$); (c) nanopillars, $L \sim W$.

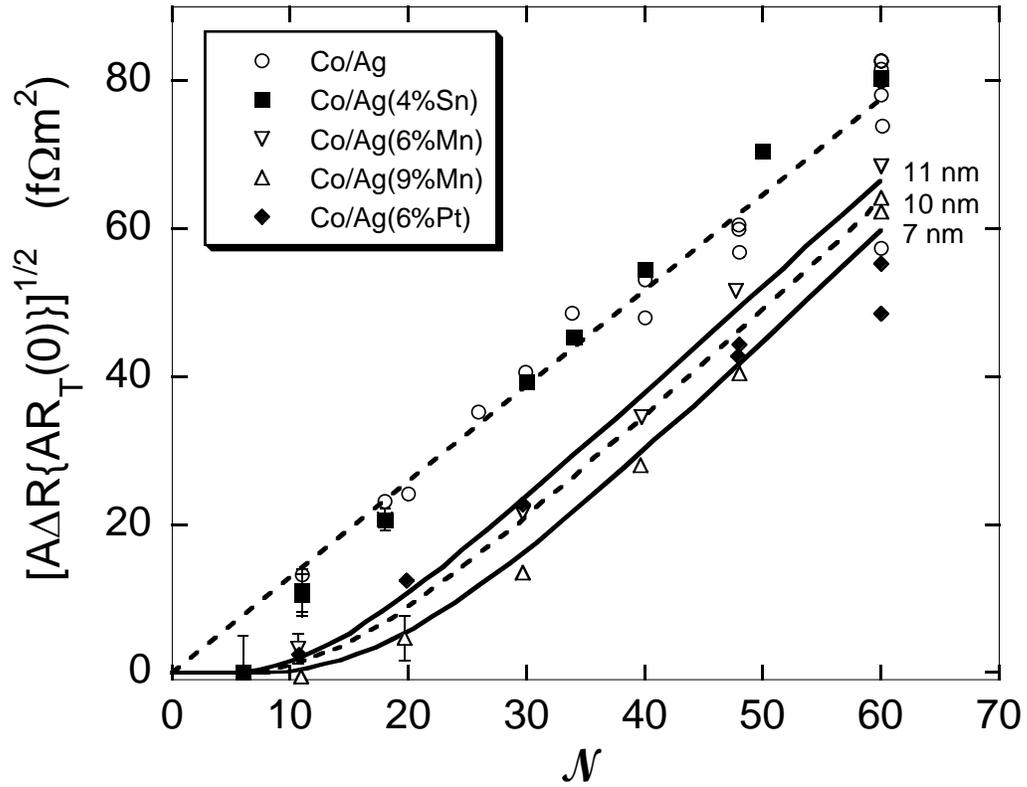

Fig. 3. $\sqrt{A\Delta R(AR(AP))}$ vs $\mathcal{N}$ for Ag and Ag-based alloys. The numbers to the right of the curves indicate $l_{sf}^N$ = 11 nm for Ag(6%Mn), 10 nm for Ag(6%Pt), and 7 nm for Ag(9%Mn). From [12]

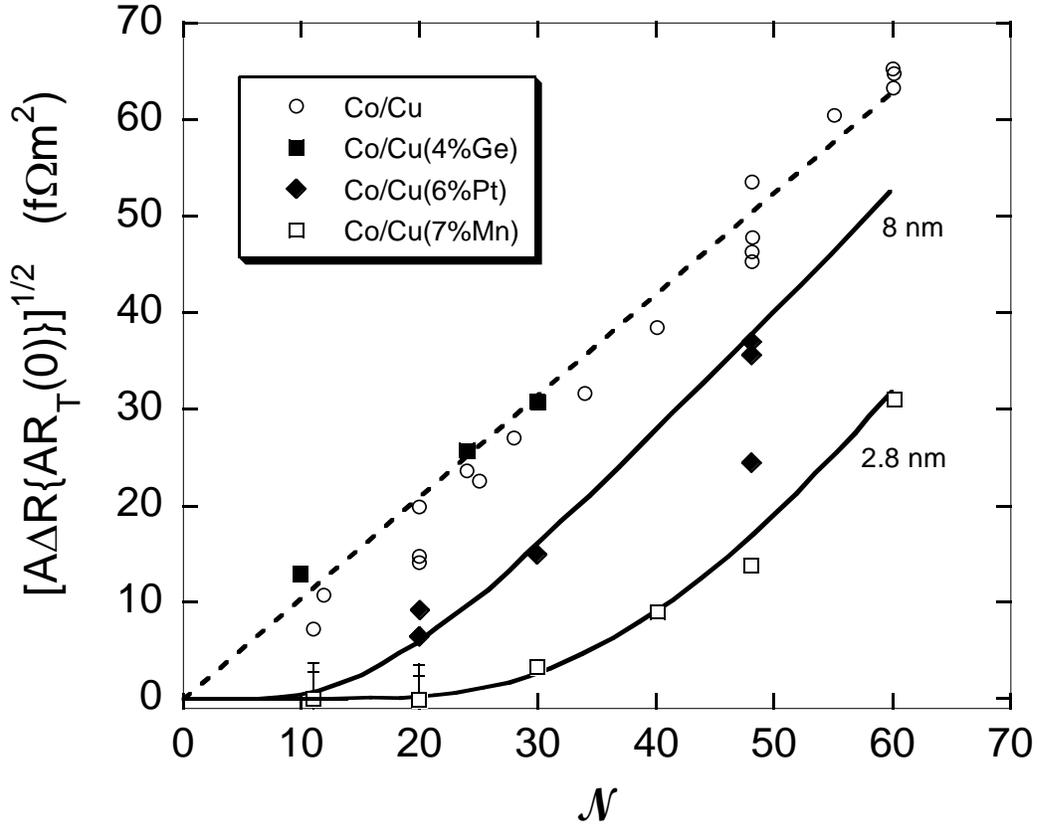

Fig. 4. $\sqrt{A\Delta R(AR(AP))}$ vs $\mathcal{N}$ for Cu and Cu-based alloys. The numbers to the right of the curves indicate $l_{sf}^N$ = 8 nm for Cu(6%Pt) and 2.8 nm for Cu(7%Mn). After [12] and [45].

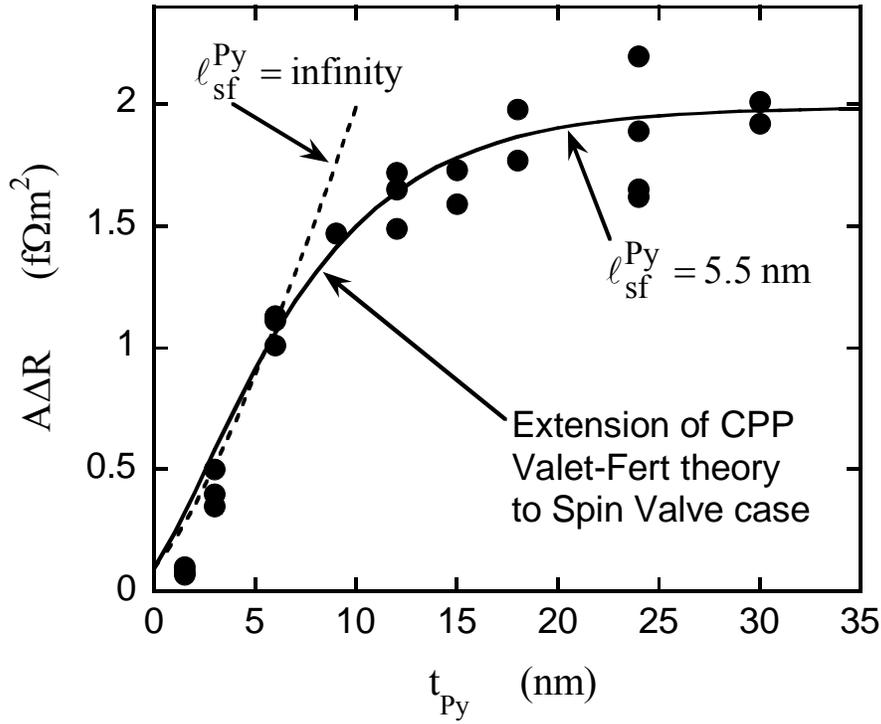

Fig. 5. AΔR vs $t_{Py}$ for Py-based EBSVs. The solid curve is a fit to VF theory with $l_{sf}^{Py}$ = 5.5 nm. The dashed curve represents the expected variation for $l_{sf}^{Py} = \infty$. Note that, for small $t_{Py}$, the solid curve lies above the dashed one, primarily because of the differences in the denominators of Eqs. 8 and 10. From [13]

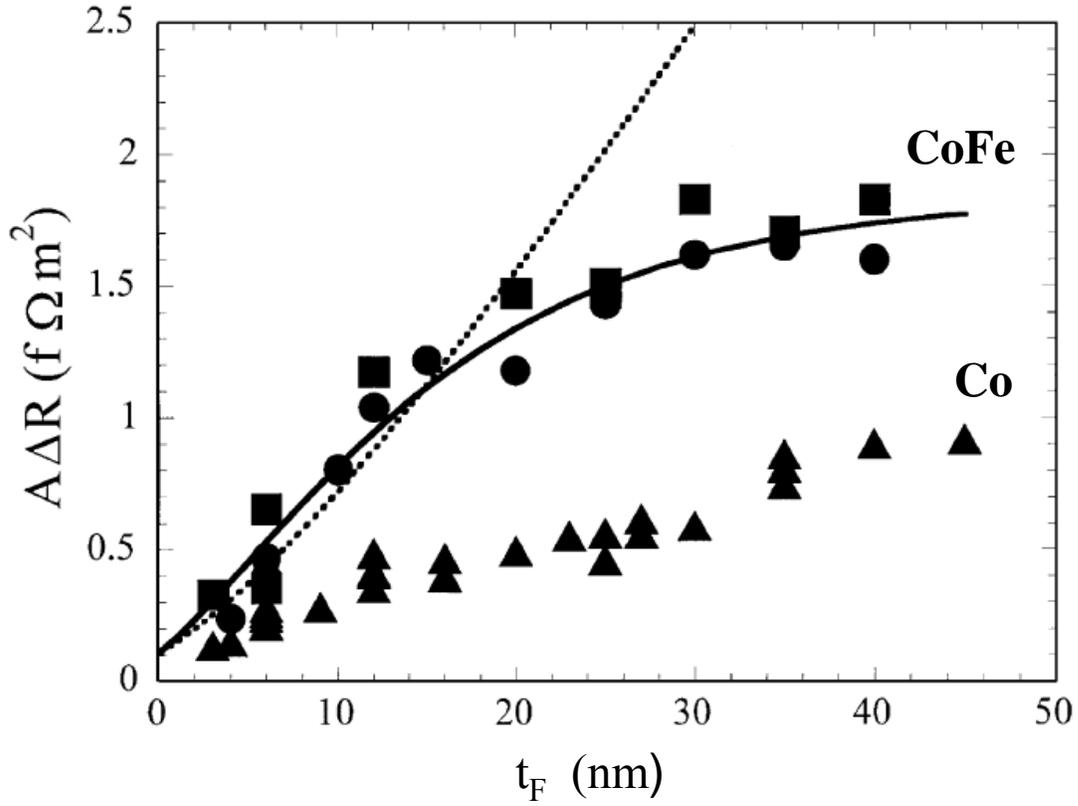

Fig. 6. AΔR vs $t_F$ for F = Co- and Co(9%Fe)-based EBSVs. The solid curve is a fit to VF theory with $l_{sf}^{CoFe}$ = 12 nm. The dashed line represents the expected variation for $l_{sf}^{CoFe} = \infty$. Note that, for small $t_{CoFe}$, the solid curve lies above the dashed one, primarily because of the differences in the denominators of Eqs. 8 and 10. From [46].

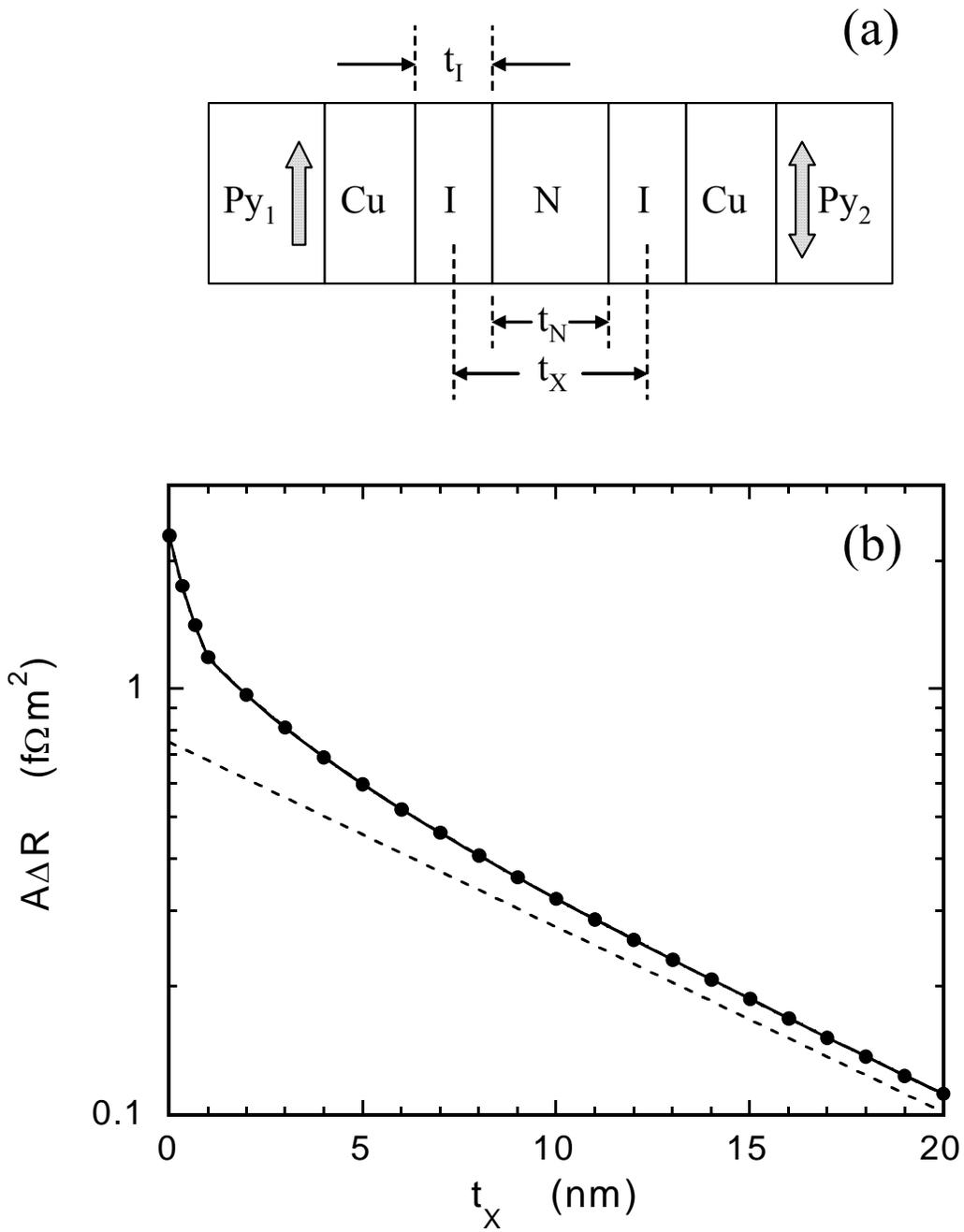

Fig. 7. (a) Schematic of an EBSV wit a single X = N-layer insert. (b) Calculation of log ($A\Delta R$) vs $t_X$ for such an EBSV. The dashed line is Eq. 11 with a constant denominator. From [48]

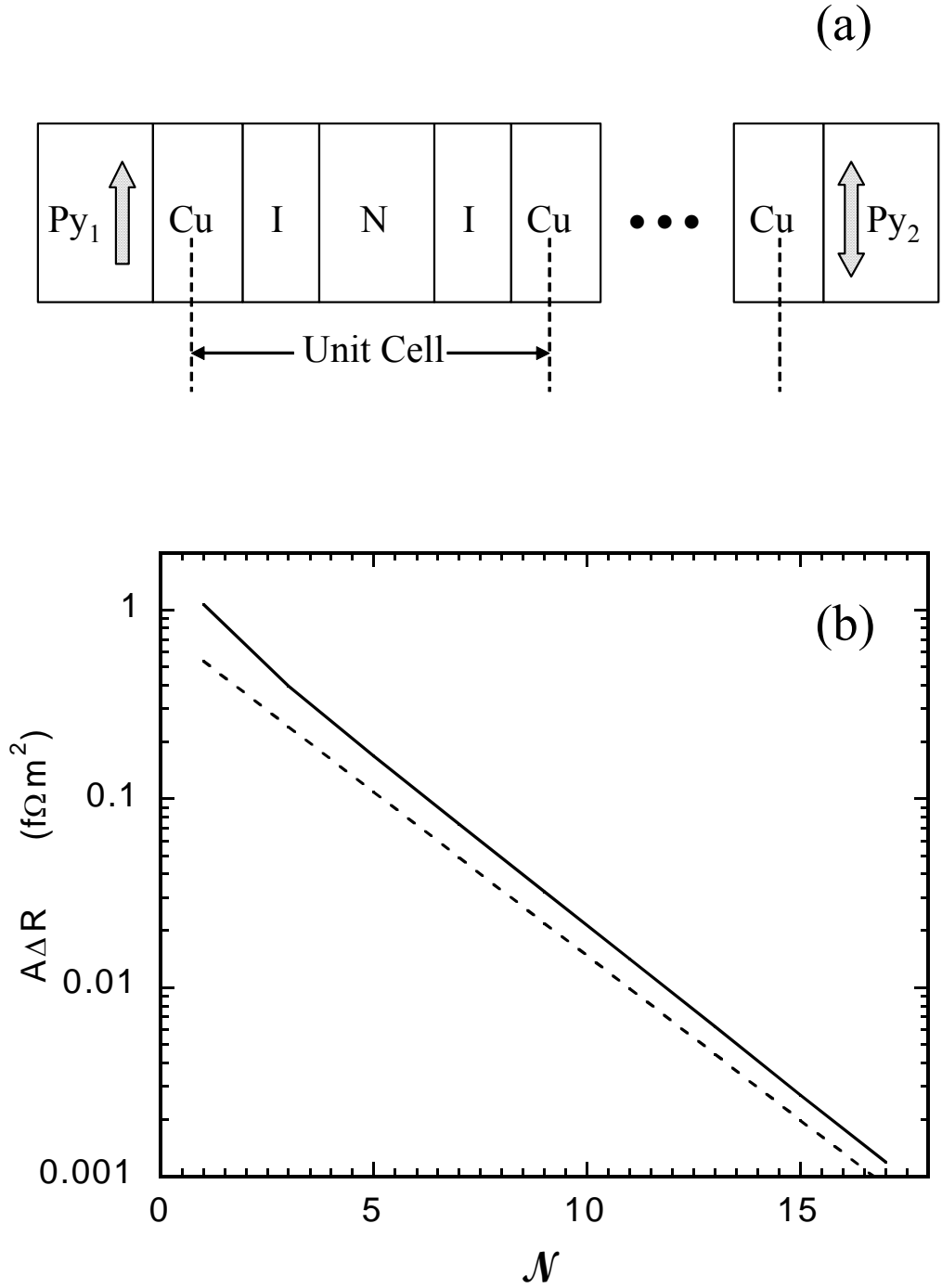

Fig. 8. (a) Schematic of an EBSV with an X = N$_1$(3)/N$_2$(1)]$_\mathcal{N}$ multilayer insert. (b) Calculated log (A$\Delta$R) vs $\mathcal{N}$ for such an insert. The dashed line is Eq. 12 with a constant denominator. To simplify, in both cases we've assumed $l_{sf}^{N1} = l_{sf}^{N2} = \infty$. From [48]

.

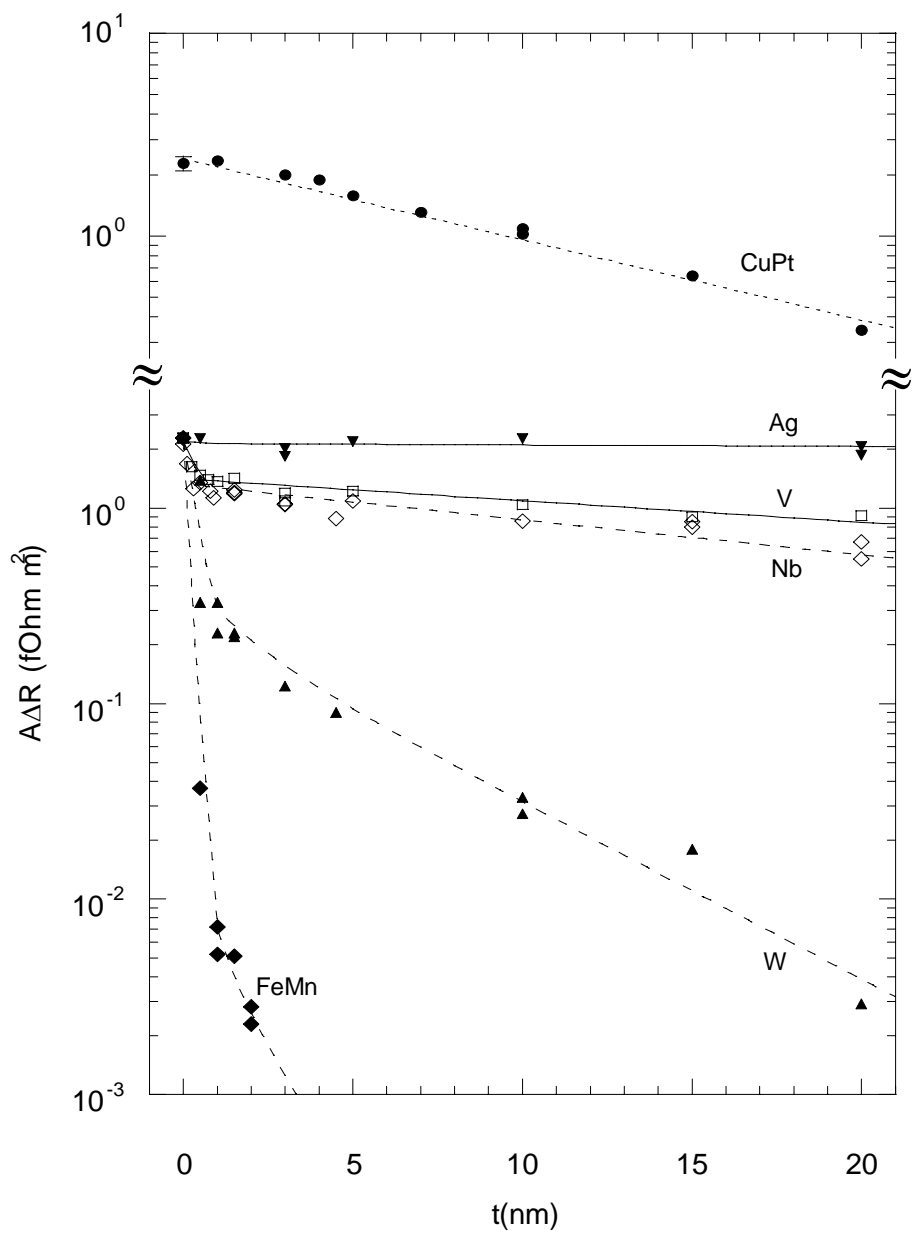

Fig. 9. log (AΔR) vs t for X = Ag, CuPt, V, Nb, W, and FeMn.  With the exception of FeMn, where the curve is just a guide to the eye, the solid and dashed curves are fits to the VF theory with the parameters in Tables II and III. From [48].

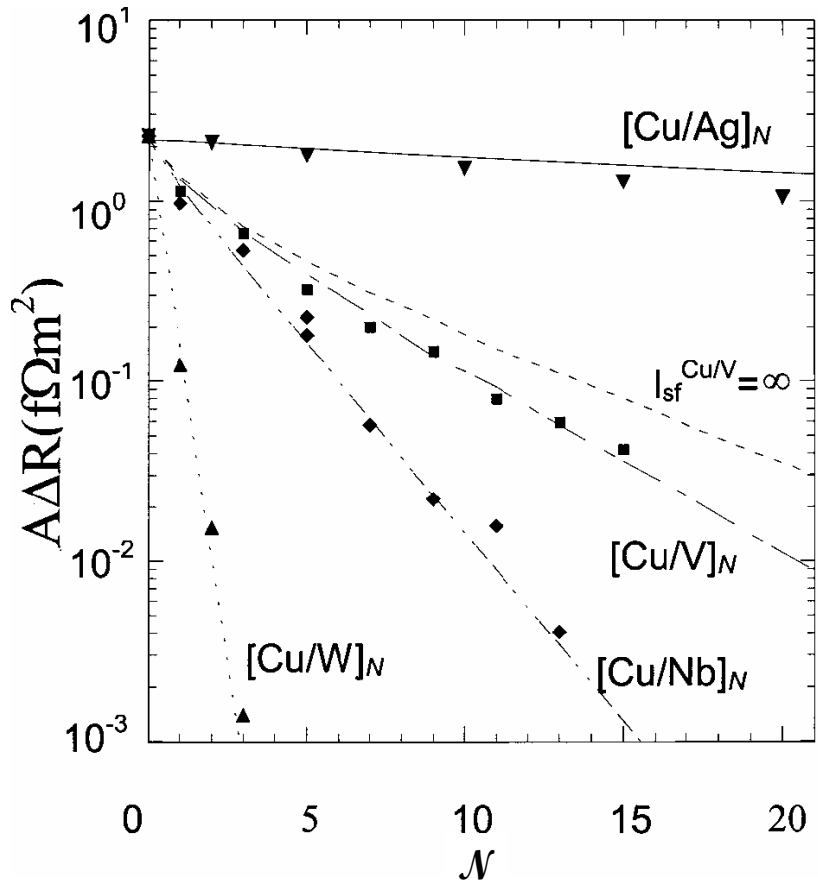

Fig. 10. log (A$\Delta$R) vs $\mathcal{N}$ for X = [Cu/Ag]$_N$, [Cu/V]$_N$, [Cu/Nb]$_N$, and [Cu/W]$_N$. The solid, broken, and dotted curves are fits using VF theory and the parameters in Tables II and III. The dashed curve indicates the expected behavior for $l_{sf}^{Cu/V} = \infty$. From [48]

38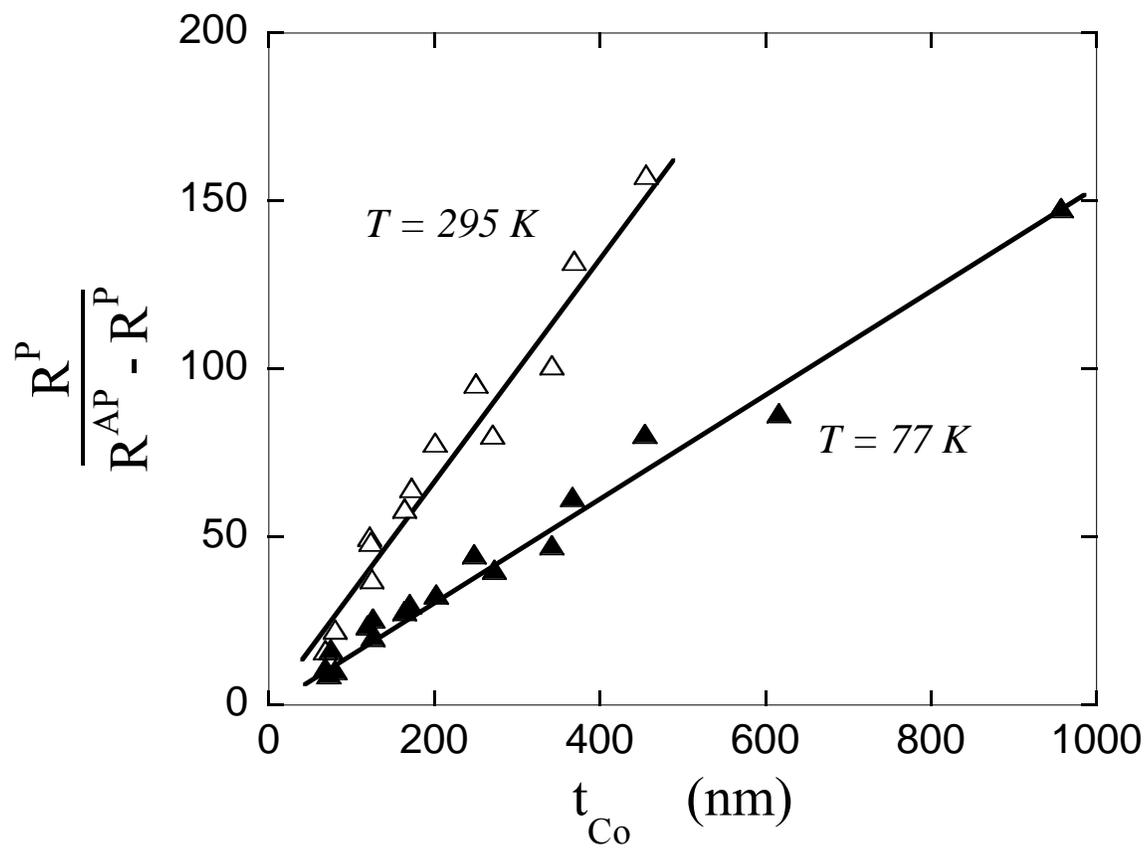

Fig. 11. R(P)/ΔR vs $t_{Co}$ at 77K and 295K. From [43].

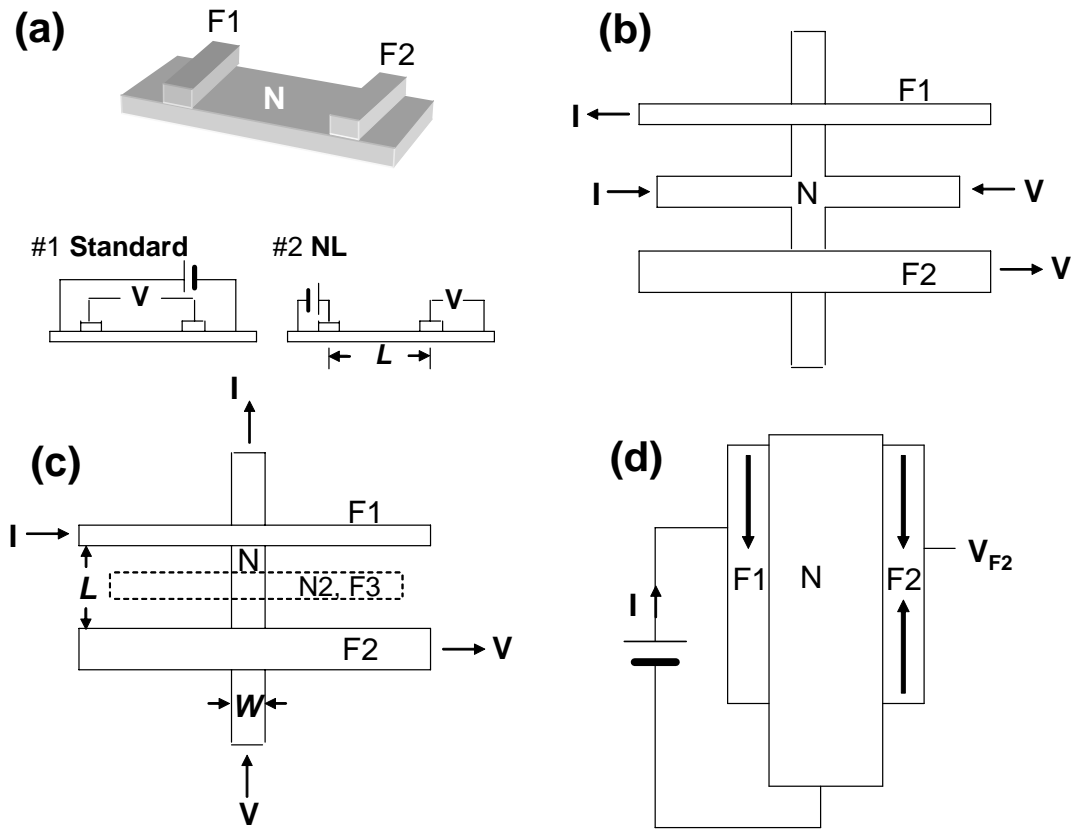

Fig. 12. Lateral (L) Geometries for Standard and Non-local (NL) Measurements. (a) Lateral spin-valve film with Standard (#1) and non-local (NL) (#2) current and voltage connections. (b) LNL-Cross (LNL/C) geometry with F1 and F2 layers of different widths. (c) LNL/+ geometry with additional N2 and/or F3 cross-strips. (d) LNL/TTD three terminal device.

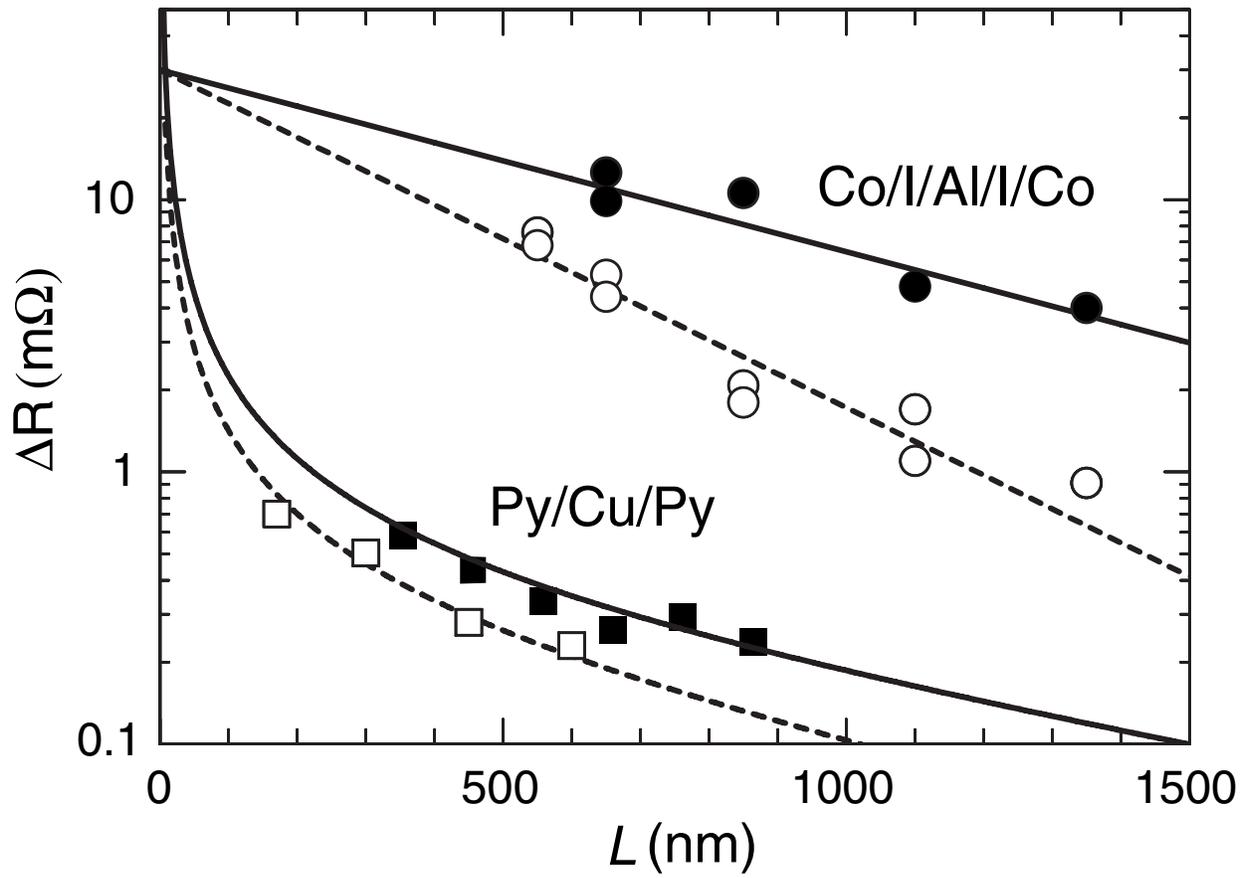

Fig. 13. ΔR vs *L* for: (a) an LNL/T Co/I/Al/I/Co sample--● = 4.2K, o = 293K, data from Jedema et al., [73]; and (b) an LNL Py/Cu/Py samples-- ■ = 4.2K, data from Garzon [83], □ = 293K, data from Kimura et al. [84]  After [82].

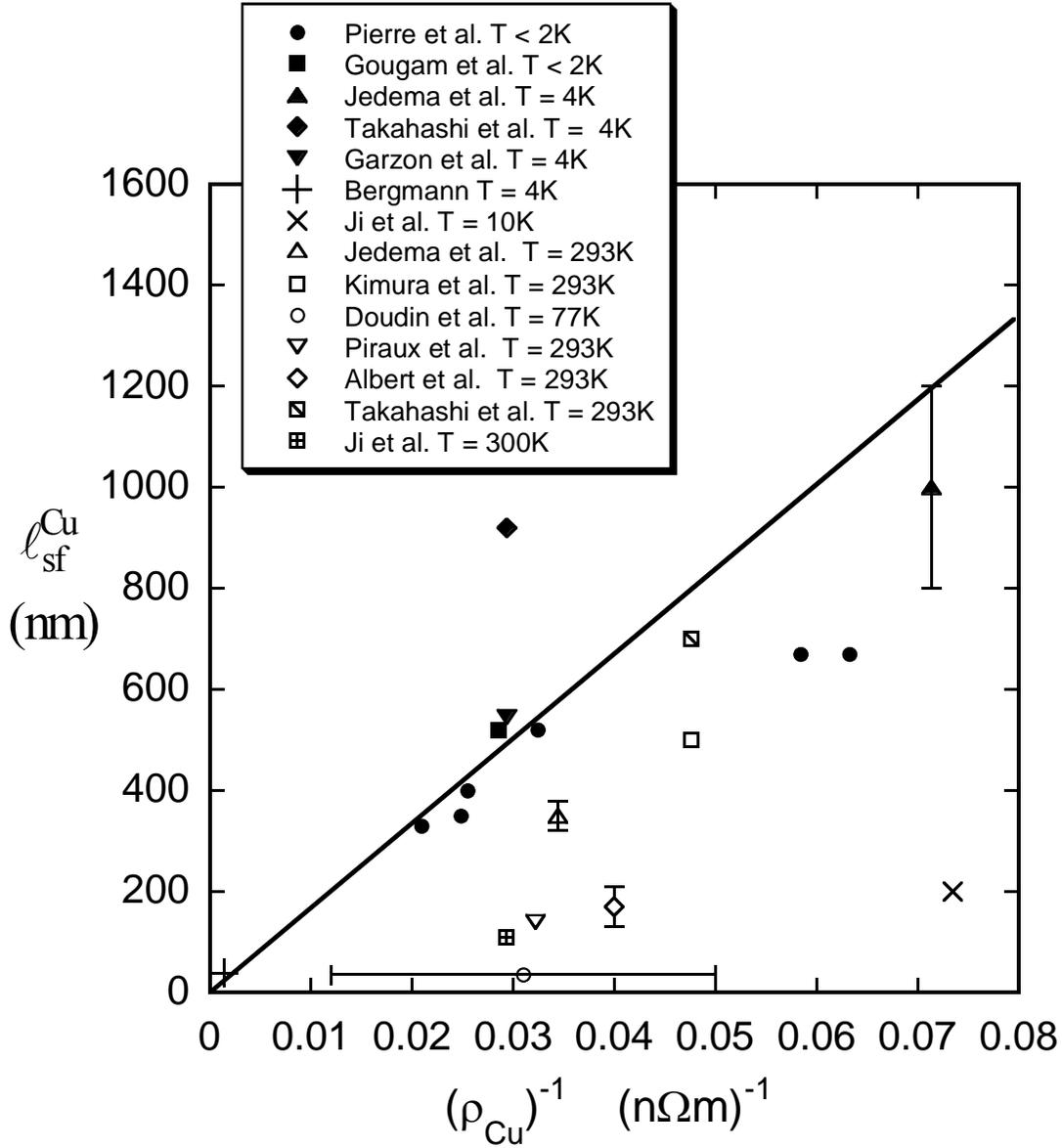

Fig. 14. $l_{sf}^{Cu}$ vs $1/\rho_{Cu}$ for Cu samples in Table II. References: Pierre [91]; Gougam [97]; Jedema [70, 77]; Takahashi [82]; Garzon [58]; Bergmann, [96], Ji [80]; Kimura [84]; Doudin [64]; Piraux [43]; Albert [65] . The line is a least-squares fit to the data for T ≤ 4.2K (filled symbols) constrained to go to (0,0) and neglecting the symbols (+ and x). Note: For pure Cu at 293K, $1/\rho_{Cu}$ = 0.060 (nΩm)$^{-1}$ [40].

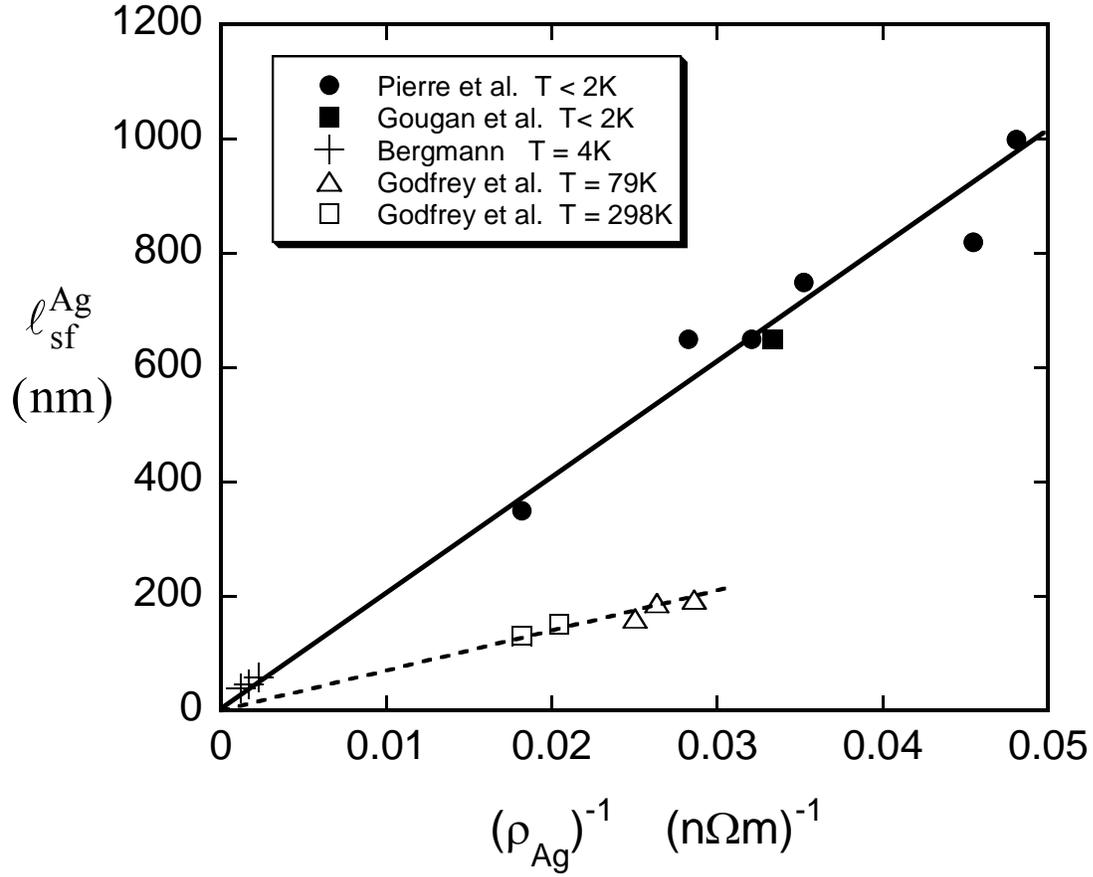

Fig. 15. $l_{sf}^{Ag}$ vs $1/\rho_{Ag}$ for Ag samples in Table II. Symbols: Pierre [91]; Gougan [97]; Bergmann [96]; Godfrey [86]. The solid line is a least-squares fit to the data for T < 2K (filled symbols) constrained to go to (0,0). The dashed line is a similar fit to the data for T = 79K and 298K. We omit from Fig. 15 the data point in Table II by Park et al. [48] which set only an extreme lower bound on $l_{sf}^{Ag}$. Note: For pure Ag at 293K, $1/\rho_{Ag} = 0.063$ $(n\Omega m)^{-1}$ [40].

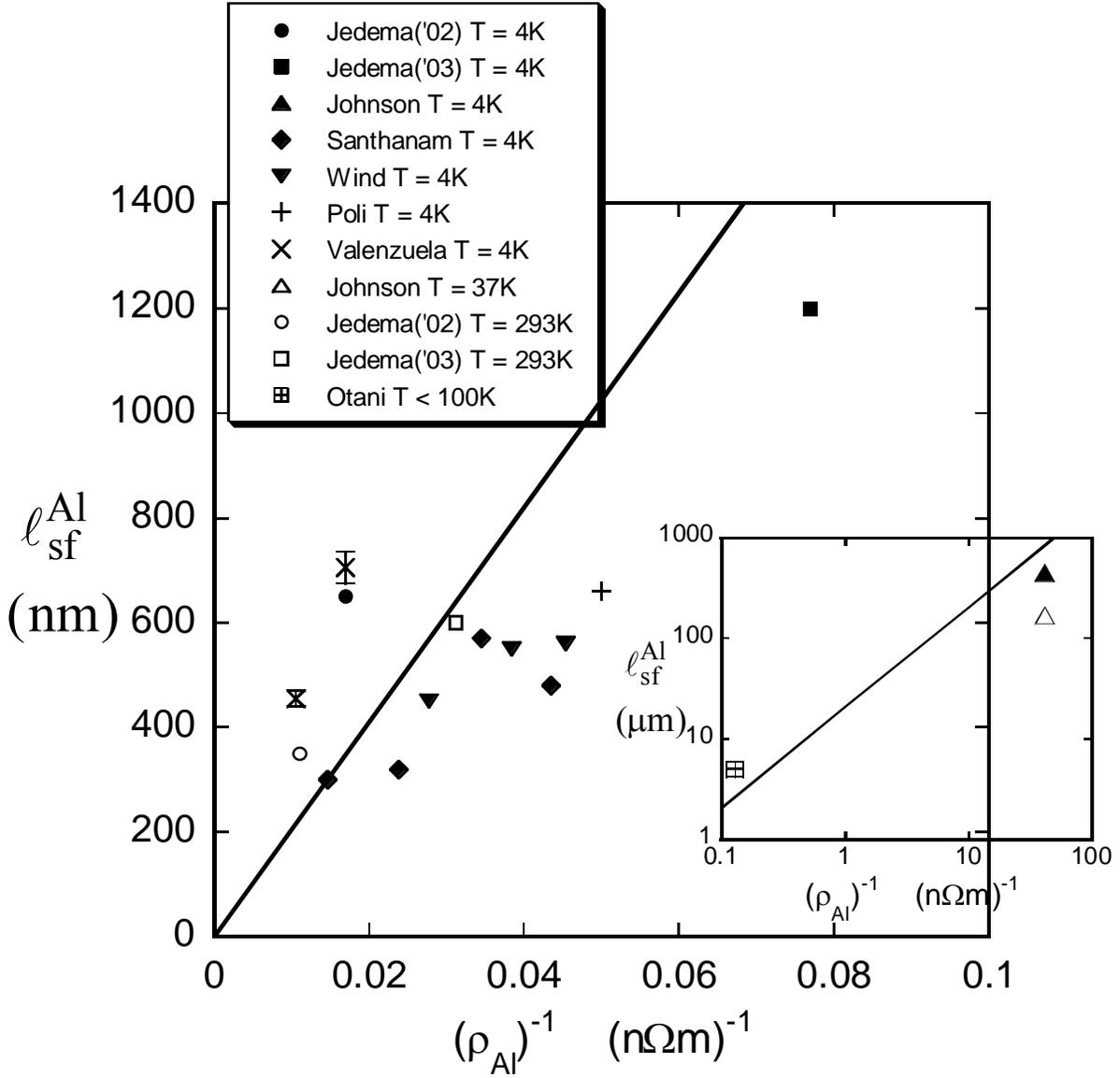

Fig. 16. $l_{sf}^{Al}$ vs $1/\rho_{Al}$ for Al samples in Table II. Symbols: Jedema (02)[73]; Jedema (03) [77]; Johnson [3]; Santhanam [24, 99]; Wind [100]; Poli [88]; Valenzuela [89]; Otani [90]. The straight line is a least-squares fit to the 4K data, constrained to go to (0,0). We use a log-log insert plot to place the higher purity samples of Johnson and Otani; the line in the insert is the same as in the main figure. Note: For pure Al at 293K, $1/\rho_{Al}$ = 0.038 $(n\Omega m)^{-1}$ [40].

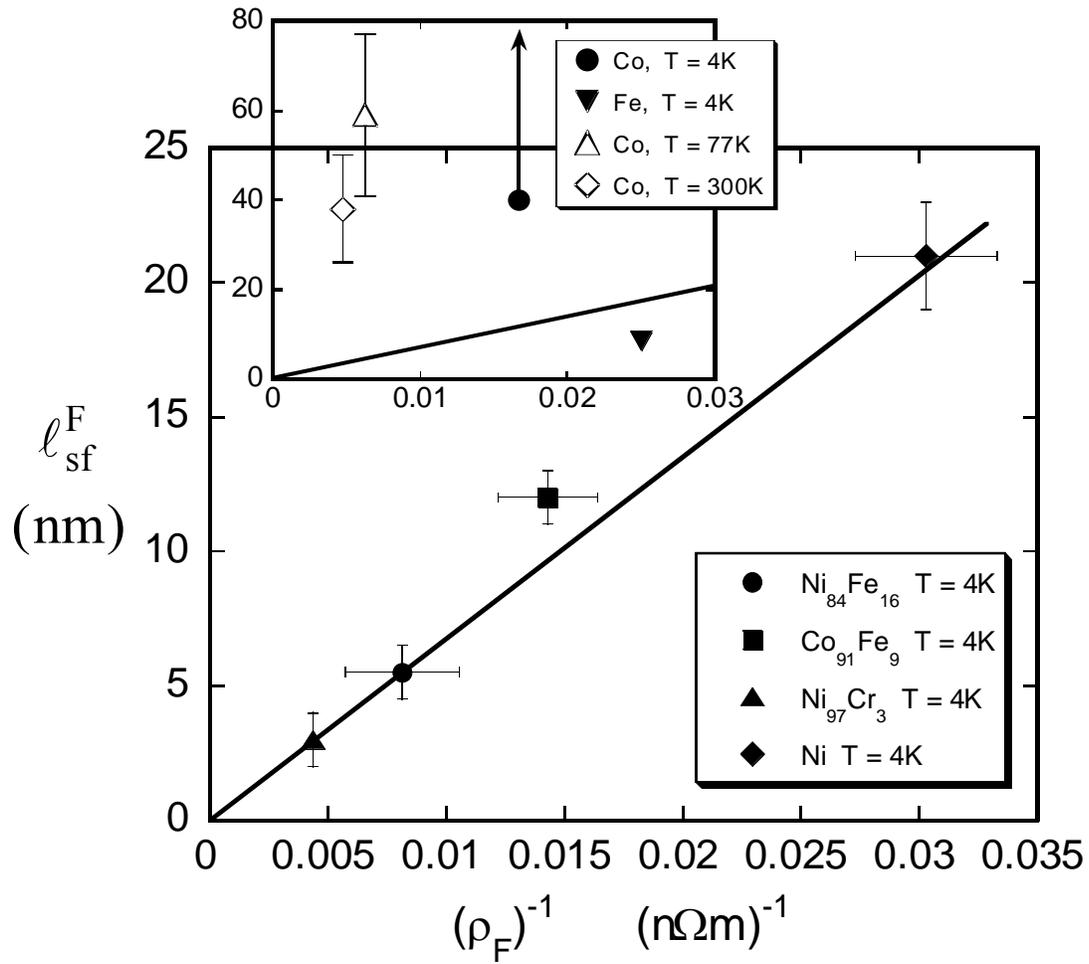

Fig. 17. $l_{sf}^F$ vs $1/\rho_F$ for CPP-MR samples in Table III. The main figure contains values for Ni and alloys, plus a best fit straight line to just those values and constrained to go to (0,0). The Insert contains this same line plus values for Fe and Co.

45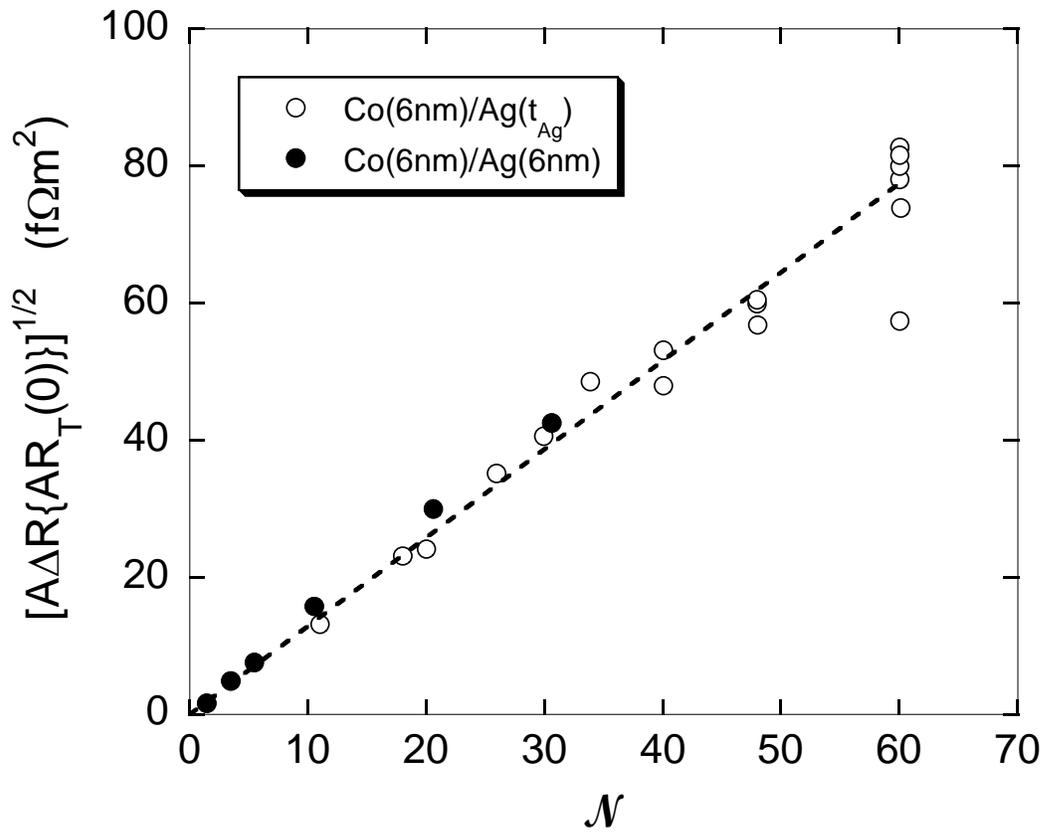

Fig. 18. $\sqrt{A\Delta R(AR(0))}$ vs $\mathcal{N}$ for Co/Ag with fixed $t_{Co}$ = 6nm comparing data for fixed $t_{Ag}$ = 6 nm and for fixed $t_T$ = 720 nm. The dashed line is a fit to the open circles passing through (0,0). After [45]